\documentclass[preprint]{aastex}
\usepackage{amsmath,amssymb}
\newcommand{\kk}{\mathbf{k}}
\newcommand{\pp}{\mathbf{p}}

\newcommand{\qq}{\mathbf{q}}
\newcommand{\xx}{\mathbf{x}}

\newcommand{\uu}{\mathbf{u}}
\newcommand{\vv}{\mathbf{v}}
\newcommand{\hh}{\mathcal{H}}

\begin{document}

\title{An Application of Wiener Hermite Expansion \\ 
        to Non-linear Evolution of Dark Matter }

\author{N. S. Sugiyama}
\affil{Astronomical Institute, Graduate School of Science, Tohoku University, Sendai 980-8578, Japan}

\author{T. Futamase }
\affil{Astronomical Institute, Graduate School of Science, Tohoku University, Sendai 980-8578, Japan}

\email{sugiyama@astr.tohoku.ac.jp}

\begin{abstract}
We apply the Wiener Hermite (WH) expansion to the non-linear evolution of the large-scale structure,
and obtain an approximate expression for the  matter power spectrum in the full order of the expansion. 
This method allows us to expand any random function in terms of an orthonormal basis in the space of random functions 
in such a way that the first order of the expansion expresses Gaussian distribution, and others are the deviation from the Gaussianity.   
It is proved that the WH expansion is mathematically equivalent to
the $\Gamma$-expansion approach in the renormalized perturbation theory (RPT).
While exponential behavior in the high-$k$ limit has been proved for the mass density and velocity fluctuations of dark matter in the RPT,
we prove the behavior again in the context of the WH expansion using the result of the standard perturbation theory (SPT). 
We propose a new approximate expression for the matter power spectrum which interpolates the low-$k$ expression
corresponding to the 1-loop level in SPT and the high-$k$ expression
obtained by taking a high-$k$ limit of the WH expansion.
The validity of our prescription is specifically verified by comparing with the 2-loop solutions of SPT.
The proposed power spectrum agrees with the result of $N$-body simulation with accuracy better than $1 \%$ or $2 \%$
in a range of the baryon acoustic oscillation scales, where the wave number is about $k$ = 0.2--0.4 $h{\rm Mpc^{-1}}$ at $z=0.5$--$3.0$. 
This accuracy is comparable to or slightly less than the ones in the closure theory, 
the fractional difference of which from the $N$-body result is within $1 \%$.
One merit of our method is that the computational time is very short because only single and double integrals are involved in our solution.
\end{abstract}


\section{Introduction}

Precise measurements of matter power spectrum in the large-scale structure are a powerful tool not only to investigate the details of the structure formation, but also to estimate the cosmological parameters.
For example, the precise measurements of the baryon acoustic oscillation (BAO) in the matter power spectrum observed by 
the Sloan Digital Sky Survey  
has emerged as a powerful tool to estimate cosmological parameters
\citep{Eisenstein/etal:1998tu,Matsubara:2004fr, Eisenstein/etal:2005,Seo:2003pu,Blake:2003rh,Glazebrook:2005mb,Shoji:2008xn,Padmanabhan:2008ag}. 
Also the observation of cosmic shear in the near future is expected to give a useful constraint on the nature of dark energy.   
Obviously, proper understanding of the observed power spectrum becomes possible
only if an accurate theoretical prediction is available  which requires a good 
understanding of the non-linear evolution of dark matter perturbation,
the relation between dark matter and baryonic matter (bias effect) and the redshift distortion effect. 
There have been various studies and much progress on the theoretical calculations of the power spectrum, 
but it is still useful and required to have more accurate theoretical treatment.  In this paper, 
we give a new approach to describe the non-linear evolution of dark matter.
It is called the ``Wiener Hermite (WH) expansion method'', 
where the stochastic nature of the cosmological density perturbation is manifestly used and the stochastic variables are expanded in terms of an orthonormal basis in the space of stochastic functions. The method was developed in the 1970s for application to turbulent theory in fluid dynamics and applied to cosmological turbulent theory by one of the authors of this paper. The method gives us a coupled equation at each perturbative order, even at the first order in such a way that the lower order quantities are modified by higher order quantities. This is totally different from the usual perturbation theory where lower order quantities are never influenced by higher order quantities. Thus, it gives us a prescription 
for the renormalization of higher order effects, and the precise meaning is described below. 
Each expansion coefficient has a clear statistical meaning; namely, the coefficients of the first, second and third terms in the expansion express the amplitude of Gaussianity, the skewness and kurtosis, respectively. Thus each term corresponds directly to an appropriate $n$-point correlation function. 

We mention here  some details on the previous approaches relation to ours.
It has been known for some time that the standard perturbation theory (SPT) of cosmological perturbation 
can be analytically solved in the Einstein-de Sitter universe in integral forms
\citep{Fry:1983cj,Goroff:1986ep,Suto:1990wf,Makino:1991rp,Jain:1993jh,Scoccimarro:1996se,Bernardeau:2001qr}.
When it is considered up to the third order in SPT (1-loop level), the analytical predictions describe the nonlinearity well at sufficiently high redshifts \citep{Jeong:2006xd,Jeong:2008rj}. 
However, the predictions are still insufficient at the observable low redshifts ($z=$ 0 -- 3), and we need to consider further non-linear effects.
Furthermore, it is computationally expensive to deal with the higher order corrections in SPT. 
Therefore, various modification of SPT  have been proposed in the past. One of the main approach is the ``Renormalized Perturbation Theory'' 
(RPT; \citep{Crocce:2005xy,Crocce:2005xz,Crocce:2007dt}), 
where the basic equations for fluid describing matter perturbation are rewritten in a convenient compact form 
in order to use a diagrammatic technique developed in quantum field theory 
\citep{Scoccimarro:1997gr}. Further modification have been considered, such as, e.g.,
the ``Closure Theory'' \citep{Taruya:2009ir,Hiramatsu:2009ki},
the ``Time Renormalization Group'' approach \citep{Pietroni:2008jx},
and the ``$\Gamma$-expansion approach'' using Multi-Point Propagators
\citep{Bernardeau:2008fa,Bernardeau:2010md,Bernardeau:2011vy,Bernardeau:2011dp}.
Many other new methods have also been studied \citep{McDonald:2006hf,Valageas:2003gm,Matarrese:2007wc}.
On the other hand, there is also an approach to the large-scale structure in the framework of the Lagrangian picture, called ``Lagrangian Resummation Theory'' (LRT; \citep{Matsubara:2007wj,Matsubara:2011ck,Matsubara:2008wx,Okamura:2011nu}).

It will be shown that our approach is mathematically equivalent to the $\Gamma$-expansion approach, but it still has the features described above and gives us a very convenient expression for the matter power spectrum described below. 
In almost all modified perturbation theories, the resummation of nonlinear effects, which means the partial summation of the infinite order 
in SPT,
is considered. This implies that any modified perturbative expansion methods should be described in the context of SPT. 
In this paper, we use only SPT, and prove again various properties of cosmological perturbations, e.g., their behavior in the small-scale limit (high-$k$ limit) 
proved in the context of RPT \citep{Crocce:2005xy,Crocce:2005xz,Bernardeau:2011vy}.

Since low-$k$ solutions can be safely computed using SPT,
the derivation of more precise solutions of cosmological perturbations by interpolating
between the 1-loop results and the high-$k$ behavior \citep{Crocce:2005xz,Bernardeau:2011dp} has been attempted.
However, some arbitrariness have remained for this prescription.
To resolve this problem, we propose a unique interpolation between the low-$k$ solutions and the high-$k$ ones by assuming that the higher order solutions in perturbation theory are well approximated by the ones in the high-$k$ limit.
Then, we precisely compute only up to 1-loop level corrections in SPT
and replace the higher order corrections with the ones calculated in the high-$k$ limit.
In this way we obtain an approximate full power spectrum, 
and the power spectrum shows 
a very good agreement with $N$-body results up to rather high-$k$ (about $\lesssim$ 0.2--0.4 $h$Mpc$^{-1}$) within 1 \% or 2 \% accuracy.

This paper is organized as follows. 
In Section 2, we first explain the stochastic properties which should be satisfied by 
the density and velocity perturbations of dark matter. 
In Section 3, we briefly review SPT, which will be used later.  Then the WH expansion technique is explained in our context in Section 4. The relationship between SPT and the WH expansion method is established there, and we also show the mathematical equivalence between the WH expansion and the $\Gamma$-expansion.  
In Section 5, we prove again 
the high-$k$ limit behavior of the cosmological perturbations in the context of SPT and propose an approximate full power spectrum,
where the lower order corrections are calculated only up to 1-loop levels in SPT and 
the higher order corrections are replaced with the high-$k$ solutions.
In Section 6, we compare our result with some other analytic predictions and $N$-body simulations.
We compute the two-point correlation function in Section 7.
We summarize our work and discuss future works in Section 8.

\section{Stochastic Nature of Cosmological Perturbations}
After decoupling, baryon and dark matter fluctuations are tightly coupled by the gravitational force,
and the evolution can then be described by pressureless fluid equations (continuity equation and Euler equation) with 
the Poisson equation for the Newton gravity. Thus our basic equations are as follows \citep{Bernardeau:2001qr}:
\begin{align}
     & \frac{\partial \rho}{\partial t} + \nabla \cdot [\rho \vv] = 0 , \notag \\
     & \frac{\partial \vv}{\partial t} + \vv \cdot \nabla \vv = -\nabla \phi , \notag \\
	 & \nabla^2 \phi + \Lambda c^2 = 4 \pi G \rho ,
		\label{ori}
\end{align}
where $\rho$, $\vv$, and $\phi$ denote the mass density, velocity and gravitational potential, respectively, 
and $\Lambda$ is the cosmological constant.

When we transform the spatial coordinates as $\xx \to a \xx$ and redefine the velocity as $\vv \equiv \dot{a}\xx + \uu$,
where $a$ is the scale factor and $\uu$ is the peculiar velocity, 
we can express Eq.~(\ref{ori}) as
\begin{equation}
\frac{\partial \rho(\tau,\xx)}{\partial \tau} + 3\hh(\tau) \rho(\tau,\xx)  +   \nabla \cdot \big[ \rho(\tau,\xx) \uu(\tau,\xx) \big] = 0 ,
\label{ori1}
\end{equation}
\begin{equation}
\frac{\partial \uu(\tau,\xx)}{\partial \tau} + \hh(\tau) \uu(\tau,\xx) + \left[ \uu \cdot \nabla \right] \uu(\tau,\xx) = - \nabla \Phi(\tau,\xx) ,
\label{ori2}
\end{equation}
\begin{equation}
		\frac{\nabla^2 \phi}{a^2} + \Lambda c^2 = 4 \pi G  \rho  , 
		\label{ori3}
\end{equation}
where the conformal time $\tau$ is defined as $a d\tau \equiv dt$, 
and the conformal Hubble parameter $\hh$ is defined as $\hh = a H$, where $H$ is the Hubble parameter.
We further defined the cosmological gravitational potential as $\Phi \equiv \phi + \frac{1}{2} \hh' x^2$.

In the standard cosmological perturbation theory,  
physical quantities are decomposed into the background part and the perturbative part. 
The background part of the mass density $\bar{\rho}$ is defined as
\begin{equation}
		\bar{\rho}(t)  \equiv \langle \rho(\xx,t) \rangle = \langle \rho(0,t) \rangle .
		\label{}
\end{equation}
where $\langle \cdots \rangle$ denotes the ensemble average and
we used the translation symmetry of the ensemble average.
On the other hand, the peculiar velocity $\uu$ has no background part because
of rotation symmetry in the average sense. 
Therefore, the perturbative part of the mass density and the peculiar velocity
has the property that their ensemble average are zero by definition:
\begin{equation}
		\langle \delta \rho \rangle = \langle \uu \rangle = 0 .
		\label{stochastic}
\end{equation}

Averaging  the above set of equations, we obtain the following background equations.
\begin{equation}
	\frac{\partial \bar{\rho}}{\partial \tau} + 3 \hh \bar{\rho} = 0,
		\label{}
\end{equation}
\begin{equation}
	\frac{\partial \hh}{\partial \tau} = -\frac{4 \pi G}{3} \bar{\rho} a^2 + \frac{1}{3}\Lambda c^2 a^2 ,
		\label{H'} 
\end{equation}
Integrating Eq.~(\ref{H'}), we find the usual Friedman equation,
\begin{equation}
		\hh^2 + c^2{\cal K} = \frac{8 \pi G}{3} a^2 \bar{\rho} + \frac{1}{3} \Lambda c^2 a^2,
		\label{}
\end{equation}
where the integral constant ${\cal K}$ is interpreted as the spatial curvature.

By subtracting the background equations from Eq.~(\ref{ori1}) (\ref{ori2}),
we find our basic equations in Fourier space as follows;
\begin{equation}
\delta^{\prime} (\tau,\kk) + \theta(\tau,\kk) =
- \int \frac{dk_1^3}{(2\pi)^3} \int \frac{dk_2^3}{(2\pi)^3} (2\pi)^3   \delta_D(\kk_1 + \kk_2 -\kk)
\alpha(\kk_1,\kk_2) \theta(\tau,\kk_1) \delta(\tau,\kk_2) ,
\label{eq1}
\end{equation}
\begin{align}
\theta^{\prime} (\tau,\kk)+ \hh \theta(\tau,\kk) + \frac{3}{2} \Omega_m \hh^2 \delta(\tau,\kk)
 = - \int \frac{dk_1^3}{(2\pi)^3} \int \frac{dk_2^3}{(2\pi)^3} 
 (2\pi)^3\delta_D(\kk_1 + \kk_2 - \kk) \beta(\kk_1,\kk_2) \theta(\tau,\kk_1) \theta(\tau,\kk_2) ,
\label{eq2}
\end{align}
where $\delta \equiv \delta \rho/\bar{\rho}$ and 
$\theta \equiv  \partial_i \uu^i$ denotes the divergence of velocity,
and $\delta_D$ denotes the three-dimensional Dirac delta distribution.
We neglected the vorticity ${\bf w} \equiv \nabla \times \uu$ because the vorticity is zero if its initial value is zero,
and even if its initial value is non-zero, it decays due to the expansion of the universe. 
The functions 
\begin{equation}
\alpha(\kk_1,\kk_2) \equiv \frac{(\kk_1 + \kk_2)\cdot \kk_1}{k_1^2} ,
\label{}
\end{equation}
\begin{equation}
\beta(\kk_1,\kk_2) \equiv  \frac{|\kk_1 + \kk_2|^2(\kk_1 \cdot \kk_2)}{2k_1^2k_2^2},
\label{}
\end{equation}
encode the nonlinearity of the evolution and satisfy the conditions 
\begin{equation}
		\alpha(\kk,-\kk) = \beta(\kk,-\kk) = 0.
		\label{albe}
\end{equation}

Note that this decomposition between background and perturbation is exact in Newtonian gravity, namely there are no backreaction terms generated from the ensemble average in Newtonian gravity, and the perturbative parts $\delta$ and $\theta$ obeying Eqs.~(\ref{eq1}) and (\ref{eq2}) 
naturally satisfy the stochastic condition in Eq.~(\ref{stochastic}): $\langle \delta \rangle = \langle \theta \rangle = 0$. More specifically,  
when the solutions for nonlinear equations are expanded perturbatively, 
the property that their ensemble averages are zero is not guaranteed in general, and 
thus a redefinition of the perturbation variables such as  $\delta \to \delta - \langle \delta \rangle$ is necessary.
In the case of cosmological SPT, the average, such as $\langle \delta (\kk) \rangle$ being proportional to $\delta_D(\kk)$,
is interpreted as the vacuum bubble diagram in the diagrammatical picture
and contributes only in infinitely large scales. This means we really need to redefine of the cosmological background.  
However, we do not need to consider this prescription for perturbative variables in Eq.~(\ref{eq1}) and (\ref{eq2}). 
This is a special feature of Newtonian gravity. It will be interesting to see how the backreaction look like in the case of general relativistic gravity in our approach.

\section{Review of Standard Perturbation Theory}
We here explain SPT very briefly which will be used later. 
The solutions in the case of an Einstein de Sitter universe, where $\Omega_m = 1$ and $\Omega_{\Lambda} = 0$, 
can be described by analytically integral forms in SPT. More explicitly, 
the solution may be written in the following perturbative form,
\begin{equation}
		\delta(z,\kk) = \sum_{n=1}^{\infty} a^n \delta_n(\kk) ,  \hspace{1cm}
		\theta(z,\kk) = - \hh \sum_{n=1}^{\infty} a^n \theta_n(\kk),
		\label{SPT}
\end{equation}
where the scale factor $a$ is a growing mode solution in the linearized theory.
When the scale factor $a$ is small, 
the series are dominated by their first term, that is, by linearized theory.
The relation between the time-independent coefficients $\delta_1(\kk)$ and $\theta_1(\kk)$ 
is shown from the continuity equation ~(\ref{eq1}) as $\delta_1(\kk) = \theta_1(\kk) \equiv \delta_L(\kk)$,
and the time-independent linear power spectrum for $\delta_L(\kk)$ is defined as
\begin{equation}
		\langle \delta_L(\kk) \delta_L(\kk') \rangle
		 = (2\pi)^3 \delta_D(\kk + \kk') P_L(k) ,
		\label{}
\end{equation}
where the amplitude of the wave vector is expressed as $k \equiv |\kk|$. 

Then, the coefficients $\delta_n(\kk)$ and $\theta_n(\kk)$ are described as follows,
\begin{equation}
		\delta_n(\kk) 
		= \int \frac{d^3q_1}{(2\pi)^3} \cdots\frac{d^3q_n}{(2\pi)^3}
		(2\pi)^3\delta_D(\kk - \qq_{1n}) F_{n}(\qq_1,\dots,\qq_n)
        \delta_L(\qq_1) \cdots \delta_L(\qq_n) ,
		\label{solution-del}
\end{equation}
\begin{equation}
		\theta_n(\kk) 
		= \int \frac{d^3q_1}{(2\pi)^3} \cdots\frac{d^3q_n}{(2\pi)^3}
		(2\pi)^3\delta_D(\kk - \qq_{1n}) G_{n}(\qq_1,\dots,\qq_n)
        \delta_L(\qq_1) \cdots \delta_L(\qq_n),
		\label{solution-the}
\end{equation}
where $\qq_{1n} \equiv \qq_1 + \qq_2 + \dots + \qq_n$ and
$F_n$ and $G_n$ are completely symmetrized functions for the wave vectors $\{\qq_1,\qq_2,\dots, \qq_n\}$.
The functions $F_n$ and $G_n$ are constructed according to the following recursion relations ($n \geq 1$) \citep{Goroff:1986ep,Bernardeau:2001qr}:
\begin{align}
		F_{n+1}(\qq_1, \dots, \qq_{n+1}) 
= \sum_{m=1}^{n} \frac{G_m(\qq_1,\dots,\qq_m)}{(2n+5)n}
& \Bigg[(2n+3)\alpha(\kk_1,\kk_2) F_{n+1-m}(\qq_{m+1},\dots,\qq_{n+1}) \notag \\
&\hspace{2cm} + 2\beta(\kk_1,\kk_2) G_{n+1-m}(\qq_{m+1},\dots,\qq_{n+1})\Bigg] ,
\label{chikujiF}
\end{align}
\begin{align}
		G_{n+1}(\qq_1, \dots, \qq_{n+1}) 
= \sum_{m=1}^{n} \frac{G_m(\qq_1,\dots,\qq_m)}{(2n+5)n}
& \Bigg[3\alpha(\kk_1,\kk_2) F_{n+1-m}(\qq_{m+1},\dots,\qq_{n+1}) \notag \\
&\hspace{2cm} + (2n+2)\beta(\kk_1,\kk_2) G_{n+1-m}(\qq_{m+1},\dots,\qq_{n+1})\Bigg],
\label{chikujiG}
\end{align}
where $\kk_1 \equiv \qq_1 + \cdots + \qq_m$,
$\kk_2 \equiv \qq_{m+1} + \cdots + \qq_{n+1}$, and $F_1 = G_1 =1$.

For $n=1$, we have
\begin{equation}
		F_2(\kk_1,\kk_2)
		= \frac{5}{7} + \frac{1}{2}\frac{\kk_1\cdot\kk_2}{k_1k_2}\left( \frac{k_1}{k_2} + \frac{k_2}{k_1} \right)
		+ \frac{2}{7} \frac{(\kk_1\cdot \kk_2)^2}{k_1^2k_2^2} ,
		\label{F2}
\end{equation}
\begin{equation}
		G_2(\kk_1,\kk_2)
		= \frac{3}{7} + \frac{1}{2}\frac{\kk_1\cdot\kk_2}{k_1k_2}\left( \frac{k_1}{k_2} + \frac{k_2}{k_1} \right)
		+ \frac{4}{7} \frac{(\kk_1\cdot \kk_2)^2}{k_1^2k_2^2} .
		\label{G2}
\end{equation}

The stochastic property $\langle \delta \rangle = \langle \theta \rangle = 0$
is specifically shown from these solutions.
When we consider the average of Eq.~(\ref{solution-del}),
we only have to consider both coefficients $\delta_n$ and $\theta_n$ 
due to the linearity of the ensemble average.
The ensemble average for $\delta_n$ is
\begin{align}
		\langle \delta_n(\kk)  \rangle
	   & = \int \frac{d^3q_1}{(2\pi)^3} \cdots\frac{d^3q_n}{(2\pi)^3}
		(2\pi)^3\delta_D(\kk - \qq_{1n}) F_{n}(\qq_1,\dots,\qq_n)
        \langle \delta_L(\qq_1) \cdots \delta_L(\qq_n) \rangle \notag \\
		& = \int \frac{d^3q_1}{(2\pi)^3} \dots\frac{d^3q_n}{(2\pi)^3}
		(2\pi)^3\delta_D(\kk - \qq_{1n}) F_{n}(\qq_1,\dots,\qq_n)
		(2\pi)^2 \delta_D(\qq_{1n})  B(\qq_1,\dots,\qq_n) ,
		\label{shiki}
\end{align}
where $B(\qq_1,\dots,\qq_n)$ is defined as
\begin{equation}
		 \langle \delta_L(\qq_1) \cdots \delta_L(\qq_n) \rangle
		 \equiv (2\pi)^3 \delta_D(\qq_{1n}) B(\qq_1, \dots , \qq_n) .
		\label{}
\end{equation}
When the function $F_n$ in Eq.~(\ref{chikujiF}) substitute into Eq.~(\ref{shiki}),
$\kk_1 + \kk_2 = \qq_{1n} = 0$
is satisfied due to the Dirac delta function in Eq.~(\ref{shiki}),
and the functions $\alpha(\kk_1,\kk_2)$ and $\beta(\kk_1,\kk_2)$ become zero from Eq.~(\ref{albe}) .
Then, it is shown that the function $F_n$ in Eq.~(\ref{shiki}) becomes zero,
and $\langle \delta \rangle = 0$.
A similar analysis can be applied to $\theta$.

Note that this stochastic property is independent of the initial conditions of $\delta_L$,
that is, the initial condition can have primordial non-Gaussianity.

\section{The Wiener Hermite Expansion}
Now, we explain our expansion method for $\delta$ and $\theta$.
Our expansion scheme should satisfy the following two properties.
First, it is known observationally and theoretically that the cosmological perturbations in the universe have a nearly Gaussian distribution. 
Thus the first order in our expansion should express the Gaussian distribution. 
Second, the expansion scheme should respect the stochastic condition of the cosmological perturbations,
$\langle \delta \rangle = \langle \theta \rangle = 0$.
Based on these two conditions, we adopt the WH expansion as our  expansion method.

\subsection{Definition of the Wiener Hermite expansion}
In the WH expansion,
the perturbation variables $\delta$ and $\theta$ are expanded as follows,
\begin{equation}
		\delta(z,\kk) = \sum_{r=1}^{\infty} 
		\int \frac{d^3p_1}{(2\pi)^3} \cdots \int \frac{d^3p_r}{(2\pi)^3}
		(2\pi)^3 \delta_D(\kk - \pp_{1r}) \delta^{(r)}_{\rm WH}(z,\pp_1,\dots,\pp_r) 
		H^{(r)}(\pp_1,\dots,\pp_r) ,
		\label{}
\end{equation}
\begin{equation}
		\theta(z,\kk) = \sum_{r=1}^{\infty} 
		\int \frac{d^3p_1}{(2\pi)^3} \cdots \int \frac{d^3p_r}{(2\pi)^3}
		(2\pi)^3 \delta_D(\kk - \pp_{1r}) \theta^{(r)}_{\rm WH}(z,\pp_1,\dots,\pp_r) 
		H^{(r)}(\pp_1,\dots,\pp_r) ,
		\label{}
\end{equation}
where the functions $H^{(r)}$ $\{r = 1,2,\dots\}$ are the stochastic variables.
The first-order $H^{(1)}$ is a white noise function which satisfies the Gaussian distribution,
\begin{equation}
	   \langle H^{(1)}(\kk_1) H^{(1)}(\kk_2) \rangle = (2\pi)^3 \delta_D(\kk_1 + \kk_2) ,
		\label{}
\end{equation}
and we further define higher order bases $H^{(r)}$ $\{r = 2,3,4, \dots \}$ in the expansion as follows:
\begin{align}
	&   H^{(2)}(\kk_1,\kk_2) \equiv H^{(1)}(\kk_1) H^{(1)}(\kk_2) - (2\pi)^3 \delta_D(\kk_1 + \kk_2) ,\notag \\
	&   H^{(3)}(\kk_1,\kk_2,\kk_3) \equiv H^{(1)}(\kk_1) H^{(1)}(\kk_2) H^{(1)}(\kk_3) -  H^{(1)}(\kk_1)(2\pi)^3\delta_D(\kk_2 + \kk_3) ,\notag \\
	&   \hspace{3cm}   -   H^{(1)}(\kk_2)(2\pi)^3\delta_D(\kk_1 + \kk_3)  -  H^{(1)}(\kk_3)(2\pi)^3\delta_D(\kk_1 + \kk_2) , \notag \\
	&    H^{(4)}(\kk_1,\kk_2,\kk_3,\kk_4) \equiv \dots \ \  .
		\label{}
\end{align}
Thus they become an orthonormal basis in the space of stochastic functions, where the ensemble average of $H^{(r)}$ is clearly zero.
\begin{equation}
		\langle H^{(r)}(\kk_1 \dots \kk_r) \rangle  = 0  , \ \ \ \mbox{ \{ $r$ = 1, 2, 3, \dots \} } .
		\label{}
\end{equation}
\begin{equation}
		\langle H^{(r)}(\kk_1 \dots \kk_r) H^{(s)}(\kk'_1 \dots \kk'_s)\rangle = 0 , \ \ \  \{ r \neq s  \}  .
 \end{equation}
Therefore, the stochastic property in Eq.~(\ref{stochastic})
is satisfied by the definition of the WH expansion.

The coefficients of the WH expansion are derived by
averaging $\delta$ and $\theta$ after 
multiplying the stochastic variable $H^{(r)}$
:
\begin{align}
		& \langle \delta(\kk) H^{(r)}(-\kk_1,\dots,-\kk_r) \rangle
		= (2\pi)^3 \delta_D(\kk-\kk_{1r}) r!  \delta^{(r)}_{\rm WH}(z,\kk_1,\dots,\kk_r) , \notag \\
		& \langle \theta(\kk) H^{(r)}(-\kk_1,\dots,-\kk_r) \rangle
		= (2\pi)^3 \delta_D(\kk-\kk_{1r}) r!  \theta^{(r)}_{\rm WH}(z,\kk_1,\dots,\kk_r) .
		\label{coef}
\end{align}

The power spectrum is described in the WH expansion by
\begin{align}
		P(z,k) & = \sum_{r=0}^{\infty} (r+1)! \int \frac{d^3p_1}{(2\pi)^3} \cdots \int \frac{d^3p_{r}}{(2\pi)^3}
		[\delta^{(r+1)}_{\rm WH}(z,\kk-\pp_{1r}, \pp_1, \dots,\pp_{r})]^2 \notag \\
		& = \sum_{r=0}^{\infty} P_{\rm WH}^{(r+1)}(z,k),
		\label{full_Power}
\end{align}
where the contribution of the $(r+1)$th order in the WH expansion to 
the power spectrum $P_{\rm WH}^{(r+1)}$ is defined as
\begin{equation}
		P_{\rm WH}^{(r+1)}(z,k) \equiv (r+1)! \int \frac{d^3p_1}{(2\pi)^3} \cdots \int \frac{d^3p_{r}}{(2\pi)^3}
		[\delta^{(r+1)}_{\rm WH}(z,\kk-\pp_{1r}, \pp_1, \dots,\pp_{r})]^2 .
		\label{Hpower}
\end{equation}

\subsection{Relation between the Standard PT and Wiener Hermite Expansion}
Since the solutions for any order of SPT are given in an Einstein de Sitter universe analytically,
any new expansion for $\delta$ and $\theta$ must be described in the context of SPT.
For the linear order, we assume 
\begin{equation}
		\delta_L(\kk) = \delta^{(1)}_1 (k) H^{(1)}(\kk) , \hspace{1cm}
		P_L(k) = [\delta^{(1)}_1(k)]^2,
		\label{}
\end{equation}
where the superscript and script indices for $\delta^{(1)}_1$ denote the order of the WH expansion and SPT, respectively. It is straightforward to include the intrinsic non-Gaussianities as higher order contributions in the WH expansion. 
From now on, we express $\delta^{(1)}_1(k) \to \delta_L(k)$.
Substituting the solutions in the SPT Eqs.~(\ref{SPT}), (\ref{solution-del}), and (\ref{solution-the}) into Eq.~(\ref{coef}),
we find the general relation of the solutions between SPT and  the WH expansion as follows:
\begin{eqnarray}
	\delta^{(r+1)}_{\rm WH}(z,\kk_1,\dots,\kk_{r+1})&=& \sum^{\infty}_{n=0} a^{2n+r+1}\delta^{(r+1)}_{2n+r+1}(\kk_1,\dots,\kk_{r+1}), \notag \\
	\theta^{(r+1)}_{\rm WH}(z,\kk_1,\dots,\kk_{r+1})&=&-\hh \sum^{\infty}_{n=0} a^{2n+r+1}\theta^{(r+1)}_{2n+r+1}(\kk_1,\dots,\kk_{r+1}),
	\label{WH-SPT-1}
\end{eqnarray}
where 
\begin{align}
		\delta^{(r+1)}_{2n+r+1}(\kk_1, \dots, \kk_{r+1}) &\equiv
		 \frac{1}{(r+1)!} \frac{(2n+r+1)!}{(2n+1)!} (2n+1)!! \delta_L(k_1) \dots \delta_L(k_{r+1}) \notag \\
        &\times\int \frac{d^3p_1}{(2\pi)^3} \cdots \frac{d^3p_n}{(2\pi)^3} 
        F_{2n+r+1}(\kk_1,\dots,\kk_{r+1},\pp_1,-\pp_1, \dots , \pp_n,-\pp_n)
        P_L(p_1) \cdots P_L(p_n) ,  \notag \\
        \theta^{(r+1)}_{2n+r+1}(\kk_1, \dots, \kk_{r+1}) &\equiv
		 \frac{1}{(r+1)!} \frac{(2n+r+1)!}{(2n+1)!} (2n+1)!! \delta_L(k_1) \dots \delta_L(k_{r+1}) \notag \\
		&\times\int \frac{d^3p_1}{(2\pi)^3} \cdots \frac{d^3p_n}{(2\pi)^3} 
        G_{2n+r+1}(\kk_1,\dots,\kk_{r+1},\pp_1,-\pp_1, \dots , \pp_n,-\pp_n)
        P_L(p_1) \cdots P_L(p_n).  \notag \\
\label{WH-SPT}
\end{align}
This expression means that the density fluctuation with order $(r+1)$ in the WH expansion 
is the sum of all the density fluctuations with order  $(2n+r+1)$ in SPT.

We can understand the relation between SPT and the WH expansion through a diagrammatical representation, 
where there is no non-dimensional coupling constant and 
the order of the loop is determined by the order of the linear power spectrum $P_L(k)$,
that is, the $n$-loop contributions contains the terms proportional to $(P_L)^{n+1}$.
Each order of the WH expansion includes 
all vertex loop contributions which come from the $\delta^{(r)}$ coefficients themselves.
The order of the vertex loop is expressed by $n$ ($n \geq 0$).
On the other hand, 
the loop contributions from irreducible diagrams, where the loop order is expressed by $r$ ($r\geq 0$),
arise only after calculating the power spectrum in Eqs.~(\ref{full_Power}) and (\ref{Hpower})
(see \citep{Bernardeau:2008fa,Bernardeau:2011dp} for details).

\subsection{Relation between the $\Gamma$-expansion and Wiener Hermite expansion}
The relation given in Eq.~(\ref{coef}) corresponds to Eq.~(17) in \citep{Bernardeau:2008fa}.
That is, the WH expansion method coincides with 
the $\Gamma$-expansion approach:
\begin{equation}
		\Gamma^{(r+1)}(z,\kk_1,\dots,\kk_{r+1}) = \delta^{(r+1)}_{\rm WH}(z,\kk_1,\dots,\kk_{r+1})/(\delta_L(k_1)\cdots\delta_L(k_{r+1})).
		\label{gamma-WH}
\end{equation}
For $r=0$ in Eqs.~(\ref{WH-SPT-1}) (\ref{WH-SPT}) and (\ref{gamma-WH}),
we have
\begin{align}
		\Gamma^{(1)}(k) &= \delta_{\rm WH}^{(1)}(z,k)/\delta_L(k) \notag \\
		&= \sum_{n=0}^{\infty}a^{2n+1}(2n+1)!! \int \frac{d^3p_1}{(2\pi)^3} \cdots \frac{d^3p_n}{(2\pi)^3}
		F_{2n+1}(\kk,\pp_1,-\pp_1, \dots , \pp_n,-\pp_n)
        P_L(p_1) \cdots P_L(p_n) .
		\label{}
\end{align}
Thus, the first order of the WH expansion, $\delta^{(1)}_{\rm WH}$, corresponds to 
the propagator in RPT.
Furthermore, 
$\delta^{(r+1)}_{r+1}$ denotes the irreducible diagrams,
and this is expressed in the $\Gamma$-expansion as
\begin{equation}
		\Gamma_{\rm tree}^{(r+1)}(z,\kk_1,\dots,\kk_{r+1}) =
		a^{r+1}\delta_{r+1}^{(r+1)}(\kk_1,\dots,\kk_{r+1})/\left( \delta_L(k_1)\dots\delta_L(k_{r+1}) \right).
		\label{delta_gamma}
\end{equation}
Note that since we focus only on SPT, we consider only growing solutions, unlike RPT.

\section{Behavior of the solutions in the high-$k$ limit}
Although the WH expansion is defined and 
the interpretation is physically and mathematically useful for understanding the nonlinear evolution of dark matter, 
we need to truncate the expansion at some order, most probably at a lower order such as $r=2$ or $3$ in order to perform the actual calculation.
However, the validity of the truncation is not guaranteed immediately.
Furthermore, the computational difficulties for the power spectrum increase very rapidly  when we increase the order of truncation.
In order to resolve these difficulties, we propose in this section an approximate semi-analytic expression for the full power spectrum including all order in the WH expansion by adopting the following assumption:
The high-order solutions in SPT become dominant in the high-$k$ limit.
Therefore, they are approximated well enough by the ones in the high-$k$ limit.
Here, we show the exponential behavior of the solutions using SPT,
which have been proved in RPT \citep{Crocce:2005xy,Crocce:2005xz,Bernardeau:2011vy}.

\subsection{Functions $F_n$ and $G_n$ in the High-$k$ Limit}
We prove that the functions $F_{r+n+1}(\kk,\kk_1,\dots,\kk_r,\pp_1,\dots,\pp_n)$ and $G_{r+n+1}(\kk,\kk_1,\dots,\kk_r,\pp_1,\dots,\pp_n)$ take
the following expression in the high-$k$ limit:
\begin{eqnarray}
		\nonumber
		F_{r+n+1}(\kk,\kk_1,\dots,\kk_r,\pp_1,\dots,\pp_n) &\to& \frac{(r+1)!}{(r+n+1)!}F_{r+1}(\kk,\kk_1,\dots,\kk_r) 
		\gamma(\pp_1)\dots \gamma(\pp_n), \\
		G_{r+n+1}(\kk,\kk_1,\dots,\kk_r,\pp_1,\dots,\pp_n) &\to& \frac{(r+1)!}{(r+n+1)!}G_{r+1}(\kk,\kk_1,\dots,\kk_r) 
		\gamma(\pp_1)\dots \gamma(\pp_n), 
		\label{theorem}
\end{eqnarray}
with 
\begin{equation}
		\gamma(\pp) \equiv \frac{\pp\cdot\kk}{p^2}.
		\label{}
\end{equation}
Here, we define the high-$k$ limit as
\begin{equation}
		|\kk| \gg \{|\pp_i|, \ i = 1,2,\dots,n \}.
		\label{high-k}
\end{equation}
From now on, we shall simplify the notations: 
$F_{r+n+1}(\kk,\kk_1,\dots,\kk_r,\pp_1,\dots,\pp_n) = F_{r+n+1}(\kk,\kk_r,\pp_n)$  and
$G_{r+n+1}(\kk,\kk_1,\dots,\kk_r,\pp_1,\dots,\pp_n) = G_{r+n+1}(\kk,\kk_r,\pp_n)$.

We prove this by induction in $n$ as follows. 
For $n=0$, Eq.~(\ref{theorem}) is clearly satisfied.
For some $n$, we assume 
\begin{eqnarray}
		\nonumber
		F_{r+n}(\kk,\kk_r,\pp_{n-1}) &\to& \frac{(r+1)!}{(r+n)!}F_{r+1}(\kk,\kk_r) \gamma(\pp_1)\dots \gamma(\pp_{n-1}), \\
		G_{r+n}(\kk,\kk_r,\pp_{n-1}) &\to& \frac{(r+1)!}{(r+n)!}G_{r+1}(\kk,\kk_r) \gamma(\pp_1)\dots \gamma(\pp_{n-1}).
		\label{nFG}
\end{eqnarray}
Then we show that the $n+1$ order satisfies the same limit. 

The functions $F_{r+n+1}(\kk,\kk_r,\pp_n)$ and $G_{r+n+1}(\kk,\kk_r,\pp_n)$ are given by Eqs.~(\ref{chikujiF}) and (\ref{chikujiG}).
Then, let us examine which terms become dominant in the high-$k$ limit in these recursion relations.
From Eq.~(\ref{high-k}), we keep only terms with scale dependence as $(k/p_1)\cdots(k/p_n)$.
Then, we have the terms proportional to $F_{r+n}(\kk,\kk_{i},\pp_n)$, $G_{r+n}(\kk,\kk_{i},\pp_n)$ for $i \leq r$, and
$\gamma(\pp_n) F_{r+n}(\kk,\kk_r,\pp_{n-1})$ and $\gamma(\pp_n) G_{r+n}(\kk,\kk_r,\pp_{n-1})$ in the high-$k$ limit.
This means that the recursion relation for $F_{r+n+1}(\kk,\kk_r,\pp_n)$ in the high-$k$ limit becomes:
\begin{align}
	&	F_{r+n+1}(\kk,\kk_r,\pp_n) \to \frac{1}{(2(r+n)+5)(r+n)} \Bigg\{\notag \\
	&(2(r+n)+3) \frac{C(r,m)}{C(n+r+1,m)}
			\Bigg[\sum_{m=1}^{r} G_m(\kk_m)\alpha(\kk_{1m},\kk+\kk_{(m+1)r}+\pp_{1n}) 
			F_{r+n+1-m}(\kk,\kk_{m+1},\dots,\kk_{r},\pp_n) \notag \\
			& \hspace{4cm} + \sum_{m=1}^{r} G_{r+n+1-m}(\kk,\kk_{m+1},\dots,\kk_{r},\pp_n)
			\alpha(\kk+\kk_{(m+1)r}+\pp_{1n},\kk_{1m}) 	F_{m}(\kk_m) \Bigg] \notag \\
			&\hspace{1.5cm} +  4\frac{C(r,m)}{C(n+r+1,m)}
 \Bigg[\sum_{m=1}^{r} G_{m}(\kk_m)\beta(\kk_{1m},\kk+\kk_{(m+1)r}+\pp_{1n}) G_{r+n+1-m}(\kk,\kk_{m+1},\dots,\kk_{r},\pp_n)\Bigg] \notag \\
	&\hspace{1.5cm} + (2(r+n)+3) \left(  \frac{n}{r+n+1} \right)
	\alpha(\pp_n,\kk+\kk_{1r}+\pp_{1(n-1)}) F_{r+n}(\kk,\kk_r,\pp_{n-1}) \notag \\
    & \hspace{1.5cm} + 4\left( \frac{n}{r+n+1} \right)
	\beta(\pp_n,\kk+\kk_{1r}+\pp_{1(n-1)}) G_{r+n}(\kk,\kk_r,\pp_{n-1})  \Bigg\},
\label{recursion1}
\end{align}
where we denote $\kk_{(m+1)r} \equiv \kk_{m+1} + \dots + \kk_r$ and $\pp_{1(n-1)} \equiv \pp_1 + \dots + \pp_{n-1}$,
and define as 
\begin{equation}
		C(n,r) \equiv \frac{n!}{r!(n-r)!}.
		\label{}
\end{equation}
Furthermore, the behavior of $\alpha$ and $\beta$ in the high-$k$ limit is
\begin{eqnarray}
		\nonumber
		\alpha(\kk_{1m},\kk+\kk_{(m+1)r}+\pp_{1n})  &\to&\alpha(\kk_{1m},\kk+\kk_{(m+1)r}),  \\
		\nonumber
		\alpha(\kk+\kk_{(m+1)r}+\pp_{1n},\kk_{1m}) &\to& \alpha(\kk+\kk_{(m+1)r},\kk_{1m}), \\
        \nonumber
		\beta(\kk_{1m},\kk+\kk_{(m+1)r}+\pp_{1n})  &\to& \beta(\kk_{1m},\kk+\kk_{(m+1)r}), \\
        \nonumber
		\alpha(\pp_n,\kk+\kk_{1r}+\pp_{1(n-1)})  &\to& \gamma(\pp_n),	\\
        \nonumber
		\beta(\pp_n,\kk+\kk_{1r}+\pp_{1(n-1)}) &\to& \frac{\gamma(\pp_n)}{2},
\end{eqnarray}
and Eq~(\ref{recursion1}) becomes
\begin{align}
	&	F_{r+n+1}(\kk,\kk_r,\pp_n) \to \frac{1}{(2(r+n)+5)(r+n)} \left(\frac{(r+1)!}{(r+n+1)!} \right) \gamma(\pp_1)\dots\gamma(\pp_n)\notag \\
	&\times \Bigg\{(2(r+n)+3)\Bigg[\sum_{m=1}^{r} G_m(\qq_m)\alpha(\tilde{\kk}_{1},\tilde{\kk}_2) F_{r+1-m}(\qq_{m+1},\dots,\qq_{r+1}) \Bigg]
	\notag \\
&\hspace{2.5cm} +  2 \Bigg[\sum_{m=1}^{r} G_m(\qq_m)\beta(\tilde{\kk}_{1},\tilde{\kk}_2) G_{r+1-m}(\qq_{m+1},\dots,\qq_{r+1})  \Bigg] \notag \\
	&\hspace{2.5cm} + (2(r+n)+3) n F_{r+1}(\kk,\kk_r)  + 2n G_{r+1}(\kk,\kk_r)  \Bigg\},
\label{recursion2}
\end{align}
where $\{\qq_1,\dots,\qq_{r+1}\} = \{ \kk,\kk_1,\dots,\kk_r\}$
and $\tilde{\kk}_1 = \qq_{1m}$, $\tilde{\kk}_2 \equiv \qq_{(m+1)(r+1)}$.

From Eq.~(\ref{chikujiF}) and Eq.~(\ref{chikujiG}),
we can show the following relations.
\begin{align}
\Bigg[\sum_{m=1}^{r} G_m(\qq_m)\alpha(\tilde{\kk}_{1},\tilde{\kk}_2) F_{r+1-m}(\qq_{m+1},\dots,\qq_{r+1}) \Bigg] 
	 =(r+1)F_{r+1}(\kk,\kk_r) - G_{r+1}(\kk,\kk_r)
		\label{FF}
\end{align}
\begin{equation}
\Bigg[\sum_{m=1}^{r} G_m(\qq_m)\beta(\tilde{\kk}_{1},\tilde{\kk}_2) G_{r+1-m}(\qq_{m+1},\dots,\qq_{r+1}) \Bigg] 
		= -\frac{1}{2} \left( 3F_{r+1}(\kk,\kk_r) - (2r+3) G_{r+1}(\kk,\kk_r) \right)
		\label{GG}
\end{equation}

Substituting Eq.~(\ref{FF}) and Eq.~(\ref{GG}) into Eq.~(\ref{recursion2}),
we can finally derive the following relation in the high-$k$ limit,
\begin{equation}
		F_{r+n+1}(\kk,\kk_r,\pp_n) \to \frac{(r+1)!}{(r+n+1)!} \gamma(\pp_1)\dots\gamma(\pp_n) F_{r+1}(\kk,\kk_r).
		\label{}
\end{equation}
Similarly, for $G_{r+n+1}$ we can show the following relation
\begin{equation}
		G_{r+n+1}(\kk,\kk_r,\pp_n) \to \frac{(r+1)!}{(r+n+1)!} \gamma(\pp_1)\dots\gamma(\pp_n) G_{r+1}(\kk,\kk_r).
		\label{}
\end{equation}

This ends the proof.

\subsection{Power Spectrum in the High-$k$ Limit}

We calculate the coefficients of the WH expansion in the high-$k$ limit,
\begin{align}
		\delta^{(r+1)}_{\rm WH}(z,\kk-\kk_{1r},\kk_r) & = \sum_{n=0}^{\infty} a^{2n+r+1} \delta_{2n+r+1}^{(r+1)}(\kk-\kk_{1r},\kk_r)  \notag \\
		& \to  	\sum_{n=0}^{\infty} a^{2n+r+1}
		 \delta_L(|\kk-\kk_{1r}|) \delta_L(k_1) \dots \delta_L(k_r) \notag \\
		 & \ \ \ \  \times F_{r+1}(\kk-\kk_{1r},\kk_r)
		\frac{1}{2^n n!}\left[ - \frac{k^2}{6\pi^2} \int dp P_L(p) \right]^n \notag \\
		& = \exp\left( -\frac{k^2 \sigma_v^2}{2} \right)	
		\delta_L(z,|\kk-\kk_{1r}|)\delta_L(z,k_1) \dots \delta_L(z,k_r) F_{r+1}(\kk-\kk_{1r},\kk_r) \notag \\
		& = \exp\left( -\frac{k^2 \sigma_v^2}{2} \right) \delta^{(r+1)}_{r+1}(z,\kk-\kk_{1r},\kk_1,\dots,\kk_r),
		\label{exp}
\end{align}
where we have used Eqs.~(\ref{WH-SPT}), (\ref{theorem}), and $(2n)!! = 2^n n!$.
We define $\sigma_v^2$ as
\begin{equation}
		\sigma_v^2 \equiv\frac{1}{6\pi^2} \int dp P_L(z,p) .
		\label{}
\end{equation}
Note that we define the $z$-dependent quantities such as $\delta_L(z,k)$ and $P_L(z,k)$ by multiplying the scale factor $a$,
but we assume that the scale factor can be replaced by the growth factor $D(z)$ 
in the general cosmological models, for which $\Omega_{\Lambda} \neq 0$:
$\delta_L(z,k) \equiv a\delta_L(k) \to D(z)\delta_L(k)$ and $P_L(z,p) \equiv a^2 P_L(p) \to D^2P_L(p)$.
This relation in Eq.~(\ref{exp}) is equivalent to Eq.~(42) in \citep{Bernardeau:2008fa}
from the relation between $\delta_{r+1}^{(r+1)}$ and $\Gamma_{\rm tree}$ in Eq.~(\ref{delta_gamma}).

Then, we describe the full power spectrum in the high-$k$ limit as
\begin{align}
		P(z,k) \to \exp\left( -k^2 \sigma_v^2 \right)
		\sum_{r=0}^{\infty}	P_{\rm irr}^{(r+1)}(z,k),
		\label{high-k power}
\end{align}
where
\begin{align}
		P_{\rm irr}^{(r+1)}(z,k) & \equiv (r+1)! \int \frac{d^3k_1}{(2\pi)^3} \dots \int \frac{d^3k_{r}}{(2\pi)^3}
		[\delta^{(r+1)}_{r+1}(z,\kk-\kk_{1r}, \kk_1, \dots,\kk_{r})]^2 \notag \\
		& = (r+1)! \int \frac{d^3k_1}{(2\pi)^3} \dots \int \frac{d^3k_{r}}{(2\pi)^3}
		\left[ F_{r+1}(\kk-\kk_{1r},\kk_r) \right]^2 P_L(z,|\kk-\kk_{1r}|) P_L(z,k_1) \dots P_L(z,k_r).
		\label{tree}
\end{align}
$P_{\rm irr}$ includes the contributions from all the irreducible diagrams.

Here, we take further high-$k$ limits in Eq.~(\ref{tree}).
\begin{equation}
		\kk \gg \{|\kk_i|,\ i = 1,\dots,r \}.
		\label{limit-k}
\end{equation}
Using Eq.~(\ref{theorem}),
we show 
\begin{align}
		P_{\rm irr}^{(r+1)} & \to (r+1)! (r+1) \int \frac{d^3k_1}{(2\pi)^3} \dots \int \frac{d^3k_{r}}{(2\pi)^3} 
                               [\delta^{(r+1)}_{r+1}(z,\kk, \kk_1, \dots,\kk_{r})]^2 \notag \\
							   & \to \frac{(k^2 \sigma_v^2)^r}{r!} P_L(z,k).
		\label{further limit}
\end{align}
Note that the limit of $\kk \gg \kk_r$ $(r \geq 1)$ is equal to
the approximation that the most effective region of $\kk_r$ to each integral is around $\kk_r \to 0$.
However, since $\kk_r$ have integral range of $0 \leq k_r < \infty$,
there necessarily exist the case of $\kk_r \sim \kk$ in the integral.
For $\kk_1 \sim \kk \gg  \kk_r$ $(\kk_r \geq 2)$, we have
\begin{align}
		P_{\rm irr}^{(r+1)} & = (r+1)! \int \frac{d^3k_1}{(2\pi)^3} \dots \int \frac{d^3k_{r}}{(2\pi)^3}
		[\delta^{(r+1)}(z,\kk-\kk_{1r}, \kk_1, \dots,\kk_{r})]^2 \notag \\
		& \to  (r+1)! \int \frac{d^3q}{(2\pi)^3}\frac{d^3k_2}{(2\pi)^3} \dots \int \frac{d^3k_{r}}{(2\pi)^3}
		[\delta^{(r+1)}(z,\qq, \kk,\kk_2, \dots,\kk_{r})]^2  ,
		\label{}
\end{align}
where we define $\kk$ as $\kk-\kk_{1r} \sim \kk-\kk_1 \equiv \qq$.
This is equal to the case of the high-$k$ limit.
The same analysis is applied to arbitrary $\kk_r$.
Therefore, the factor $(r+1)$ is multiplied in Eq.~(\ref{further limit}).

From Eq.~(\ref{further limit}) and Eq.~(\ref{high-k power}),
we finally derive
\begin{align}
		P(z,k) &\to P_L(z,k)\exp(-k^2 \sigma_v^2)\sum_{r=0}^{\infty}\frac{(k^2 \sigma_v^2)^r}{r!} \notag \\
			   & = P_L(z,k).
		\label{P_L}
\end{align}
Surprisingly, the solutions in the high-$k$ limit cancel each other, and the full power spectrum reduces to the linear power spectrum.
This fact is well known in the 1-loop level of SPT~\citep{Makino:1991rp},
but it is interesting that this cancellation also applies for the dominant terms in the high-$k$ limit in the full power spectrum. 
Of course, it is really not that the full power spectrum becomes the linear power spectrum in the high-$k$ limit,
because we have chosen only dominant terms in the high-$k$ limit in the proof of Eq.~(\ref{theorem}) and Eq.~(\ref{further limit}),
and the subleading terms can also affect the full power spectrum even in the high-$k$ limit. 
This result implies that the nonlinear corrections for the power spectrum generally tend to cancel each other,
and result in small corrections as specifically known for 1 and 2-loop cases of SPT.

\subsection{Approximate Full Power Spectrum}
We now propose an appropriate interpolation between the low-$k$ and high-$k$ solutions.
The low-$k$-solutions are the 1-loop solutions in SPT, 
while the high-$k$ solutions are given by Eq.~(\ref{high-k power}) derived in the previous subsection.

In order to have an expression applicable for the case of $r=0$,
we use Eq.~(\ref{theorem}) in the following
\begin{equation}
	F_{2n+1}(\kk,\pp_1,-\pp_1,\dots,\pp_n,-\pp_n)
	\to \frac{3!}{(2n+1)!} F_3(\kk,\pp,-\pp) \gamma(\pp_2)\gamma(-\pp_2)\dots\gamma(\pp_n)\gamma(-\pp_n).
		\label{wh1ap}
\end{equation}
Then, for $n \geq 1$, we show
\begin{align}
		\delta_{2n+1}^{(1)}(k)
		&\to \frac{2(2n+1)!!}{(2n+1)!} \bigg[\delta_L(k) 3\int \frac{d^3p_1}{(2\pi)^3} F_3(\kk,\pp_1,-\pp_1)P_L(p_1) \bigg]
		\left[ \int \frac{d^3p}{(2\pi)^3} \gamma(\pp) \gamma(-\pp) P_L(p)\right]^{n-1} \notag \\
		& =\delta^{(1)}_3(k)\frac{2}{2^nn!}\left[- \frac{k^2}{6\pi^2} \int dp P_L(p) \right]^{n-1},
		\label{}
\end{align}
where we have denoted the 1-loop correction term in SPT as
\begin{equation}
		\delta_3^{(1)}(k) = 3\delta_L(k) \int \frac{d^3p}{(2\pi)^3} F_3(\kk,\pp,-\pp) P_L(p).
		\label{}
\end{equation}

Then, we derive the approximate solution of $\delta^{(1)}_{\rm WH}$ as
\begin{align}
		\delta^{(1)}_{\rm WH}(z,k) & = \sum_{n=0}^{\infty} D^{2n+1} \delta_{2n+1}^{(1)}(k) \notag \\
		& \to \delta_L(z,k) + \delta_3^{(1)}(z,k)
		\left( \frac{2}{-k^2 \sigma_v^2} \right)\sum_{n=1}^{\infty} \frac{1}{n!}\left( -\frac{k^2 \sigma_v^2}{2} \right)^{n}  \notag \\
		& =  \delta_L(z,k) - \frac{2\delta_3^{(1)}(z,k)}{k^2 \sigma_v^2} \left[ \exp\left(-\frac{k^2 \sigma_v^2}{2}\right) -1 \right].
		\label{h1}
\end{align}
where we have used the general growth factor $D$ instead of the scale factor $a$.
The contribution to the power spectrum $P_{\rm WH}^{(1)}$ in Eq.~(\ref{Hpower}) is 
\begin{equation}
		P_{\rm WH}^{(1)}(z,k) \to \left[ 
		 \delta_L(z,k) - \frac{2\delta_3^{(1)}(z,k)}{k^2 \sigma_v^2} \left( \exp\left(-\frac{k^2 \sigma_v^2}{2}\right) -1 \right) \right]^2.
		\label{H1}
\end{equation}
For low-$k$,
we can expand the exponential term as $e^{-k^2 \sigma^2_v/2} \sim 1 - k^2 \sigma^2_v/2$,
and this lead the 1-loop correction,
\begin{equation}
		\delta^{(1)}_{\rm WH}(z,k) \to \delta_L(z,k) + \delta_3^{(1)}(z,k).
		\label{}
\end{equation}
While, for the high-$k$ limit, $\delta^{(1)}_{\rm WH}$ becomes coincident with $\delta_Le^{-k^2 \sigma_v^2/2}$
due to the good convergence of $\delta_L + 2 \delta_3^{(1)}/(k^2 \sigma_v^2) \to 0$.

Next, for $r \geq 1$, we use the approximation of Eq.~(\ref{high-k power}).
Here, we further approximate $P_{\rm irr}^{(r+1)}$ because 
it is expensive to compute the terms in the case of $(r > 2)$ due to their large multiple integrals.
Using the following approximation from Eq.~(\ref{theorem}),
\begin{equation}
		F_{r+1}(\kk-\kk_{1r},\kk_1,\dots,\kk_r)
		\to \frac{2!}{(r+1)!} \gamma(\kk_2) \dots, \gamma(\kk_r) F_2(\kk-\kk_1, \kk_1),
		\label{whrap}
\end{equation}
we derive the approximate solution of $P_{\rm WH}^{(r+1)}$ as,
\begin{align}
		P_{\rm WH}^{(r+1)}(z,k) & \to\exp\left( -k^2 \sigma_v^2 \right) (r+1)! \int \frac{d^3k_1}{(2\pi)^3} \dots \frac{d^3k_r}{(2\pi)^3} 
			\left[  \delta_{r+1}^{(r+1)}(z,\kk-\kk_{1r},\kk_1,\dots,\kk_r) \right]^2 \notag \\
			& \to \exp\left( -k^2 \sigma_v^2 \right) (r+1)! \frac{2!}{(r+1)!} \frac{2!}{(r+1)!} \left( \frac{r+1}{2} \right)
			\int \frac{d^3k_1}{(2\pi)^3}
			\left[  \delta_{2}^{(2)}(z,\kk-\kk_1,\kk_1) \right]^2 (k^2 \sigma_v^2)^{r-1}\notag \\
			& \to \exp\left( -k^2 \sigma_v^2 \right)  \frac{P_{22}(z,k)}{k^2 \sigma_v^2} \frac{1}{r!}(k^2 \sigma_v^2)^{r},
		\label{WH r>1}
\end{align}
where we multiply in the factor $(r+1)/2$ for the same reason as in Eq.~(\ref{further limit}).
We have denoted another 1-loop correction term in the SPT as,
\begin{equation}
		P_{22}(z,k) = 2 \int \frac{d^3p}{(2\pi)^3} \left[ F_2(\kk-\pp,\pp) \right]^2 P_L(z,|\kk-\pp|) P_L(z,p).
		\label{}
\end{equation}
Indeed, if for $r=1$ the exponential factor is expanded as $e^{-k^2 \sigma_v^2} \sim 1 $,
the 1-loop correction, $P_{22}$, is reproduced in Eq.~(\ref{WH r>1}).

Finally, we achieve the approximate full power spectrum,
\begin{eqnarray}
		P(z,k) & =& \sum_{r=0}^{\infty} P_{\rm WH}^{(r+1)}(z,k)  \notag \\
		                  & \to&\left[ \delta_L(z,k) - \frac{2\delta_3^{(1)}(z,k)}{k^2 \sigma_v^2}
						  \left( \exp\left(-\frac{k^2 \sigma_v^2}{2}\right) -1 \right) \right]^2 
				 + \exp\left( -k^2 \sigma_v^2 \right)  \frac{P_{22}(z,k)}{k^2 \sigma_v^2} \sum_{r=1}^{\infty}\frac{1}{r!}(k^2 \sigma_v^2)^{r},  
						  \notag \\
		P_{\rm AF}(z,k)	 & \equiv &\left[ \delta_L(z,k) - \frac{2\delta_3^{(1)}(z,k)}{k^2 \sigma_v^2}
				 \left( \exp\left(-\frac{k^2 \sigma_v^2}{2}\right) -1 \right) \right]^2 
				 + \frac{P_{22}(z,k)}{k^2 \sigma_v^2}\left[ 1 - \exp\left( -k^2 \sigma_v^2 \right) \right].
	    \label{result1}
\end{eqnarray}
This is the main result of this paper.
This gives an appropriate interpolation between the low-$k$ solutions and the high-$k$ limit ones.
We can derive the approximate solutions of each order in the WH expansion
and SPT using approximation such as Eq.~(\ref{wh1ap}) and Eq.~(\ref{whrap}),
and therefore we call this method as the ``Approximate Full Wiener Hermite (AFWH)'' expansion method
or ``Approximate Full Perturbation Theory''.
We can easily compute the solution of Eq.~(\ref{result1}) numerically, because the solution has only single or double integrals.

\section{Comparison with Other Analytic Predictions and $N$-body Simulations}

We compare the approximate full power spectrum, $P_{\rm ap}$, in Eq.~(\ref{result1}) with 
some other analytic predictions and $N$-body simulations.
We mainly use $N$-body simulations presented in \citep{Taruya:2009ir},
but in Sec.~\ref{closure} we also use other $N$-body results with higher resolutions in \citep{Valageas:2011up}.
It is plotted for the cosmological models with the 
{\it Wilkinson Microwave Anisotropy Probe} ($WMAP$) five year \citep{Komatsu:2008hk}.
cosmological parameters: 
$\Omega_m = 0.279$, $\Omega_{\Lambda} = 0.721$, $\Omega_b = 0.046$, $h = 0.701$, $n_s = 0.96$, $\sigma_8 = 0.817$.

The $N$-body simulation data with low-resolutions and high-resolutions in \citep{Taruya:2009ir} and \citep{Valageas:2011up}
were created by a public $N$-body code {\it GADGET2} \citep{Springel:2005mi}.
Their initial conditions were generated by the {\it 2LPT} code \citep{Crocce:2006ve} at $z_{\rm ini} = 31$ and $z_{\rm ini}=99$, respectively.
While the $N$-body simulations with low-resolutions 
were computed with a cubic box of size $1 h^{-1} {\rm Gpc}$ containing $512^3$ particles,
the $N$-body results with high resolutions, called L11-N11 and L12-N11,
contain $2048^{3}$ particles and were computed by combining the results with different box sizes, 
2,048 $h^{-1}{\rm Mpc}$ and 4096 $h^{-1}{\rm Mpc}$.
Although the realization of the simulations with high resolutions is only 1,
the simulations with low resolutions have the output data of 30 independent realizations,
and consider the correction of the finite-mode sampling by \citep{Nishimichi:2008ry}.
Therefore, the size of each error bar for $N$-body results with low-resolutions becomes hard to see visually.
For details of the $N$-body simulation used in this paper,
see Taruya et al. \citep{Taruya:2009ir} and Okamura et al. \citep{Okamura:2011nu}.

\subsection{Comparison with Other Analytical Predictions: 1-loop Level}

\begin{figure}[!t]
				\begin{center}
						\epsscale{0.75}
						\plotone{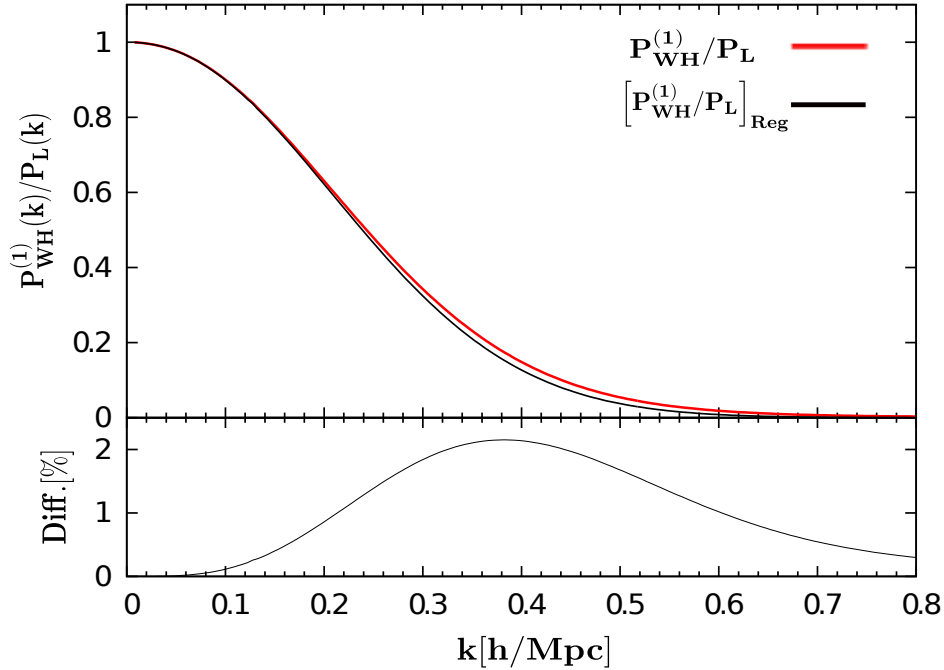}
				\end{center}
				\caption{ Comparison between $P_{\rm WH}^{(1)}$ in Eq.~(\ref{H1}) (red line) and 
				$[ P_{\rm WH}^{(1)} ]_{\rm Reg}$ in Eq.~(\ref{Hreg1}) (black line) for $P_{\rm WH}^{(1)}/P_L$ at $z=1$.
				The fractional difference, $( P_{\rm WH}^{(1)} - [ P_{\rm WH}^{(1)} ]_{\rm Reg} )/ P_L$,
				is also plotted.
				 }
		\label{fig:propagator}
\end{figure}

\cite{Bernardeau:2011dp}
proposed a simple scheme to interpolate between the low-$k$ and high-$k$ solutions, based on the $\Gamma$-expansion method.
In the scheme, 
the solutions are regularized 
so that the low-$k$ solutions become the ones in SPT and the high-$k$ solutions become Eq.~(\ref{high-k power}).
We call this scheme ``regularized $\Gamma$-expansion''.
From Eqs.~(20) (51) in \citep{Bernardeau:2011dp,Taruya:2012ut},
its 1-loop solution for the power spectrum is given by
\begin{equation}
		P_{\rm Reg} = \exp(-k^2 \sigma_v^2) \left[ \left( \delta_L + \delta_3 + \frac{k^2 \sigma_v^2}{2} \delta_L \right)^2 + P_{22} \right].
		\label{Reg}
\end{equation}

Since the WH expansion and the $\Gamma$-expansion are completely equivalent to each other,
we can understand this solution as the truncation up to the second order in the WH expansion
and write $P_{\rm WH}^{(1)}$ using the regularized $\Gamma$-expansion as
\begin{equation}
		\left[P_{\rm WH}^{(1)} \right]_{\rm Reg}
		= \exp\left( -k^2 \sigma_v^2 \right) \left[ \delta_L + \delta_3 + \frac{k^2 \sigma_v^2}{2} \delta_L  \right]^2.
		\label{Hreg1}
\end{equation}
The difference between Eq.~(\ref{H1}) and Eq.~(\ref{Hreg1}) is manner of interpolating between the solutions.
Since the regularized $\Gamma$-expansion method is a heuristic scheme,
we have used the approximation of Eq.~(\ref{wh1ap}).

When we ignore the contributions of $P_{\rm WH}^{(r+1)} (r > 2)$,
we derive the solution of the approximate WH expansion corresponding to the regularized $\Gamma$-expansion in Eq.~(\ref{Reg}) as
\begin{equation}
		P_{\rm WH}^{(1)} + P_{\rm WH}^{(2)} =	\left[ \delta_L(z,k) - \frac{2\delta_3^{(1)}(z,k)}{k^2 \sigma_v^2}
				 \left( \exp\left(-\frac{k^2 \sigma_v^2}{2}\right) -1 \right) \right]^2 
				 + \exp(-k^2 \sigma_v^2)P_{22}(z,k). 
		\label{wh12}
\end{equation}

We plot these two solutions, $P_{\rm WH}^{(1)}/P_L$ and $[P_{\rm WH}^{(1)}]_{\rm Reg}/P_L$, and their fractional difference, 
$(P_{\rm WH}^{(1)}- [ P_{\rm WH}^{(1)} ]_{\rm Reg} )/P_{L}$, at $z=1$ in Fig.~\ref{fig:propagator}.
On BAO scales ($\sim 0.2 h{\rm Mpc}^{-1}$), the fractional difference is within 1 \%.
At high $k$, our solution becomes slightly larger than the regularized $\Gamma$-expansion.
However, there is no means of investigating which results are more accurate in detail,
because on such scales the amplitude of $P_{\rm WH}^{(1)}$ decays enough due to the exponential factor 
to not largely contribute to the full power spectrum.

\begin{figure}[!t]
				\begin{center}
						\epsscale{0.75}
						\plotone{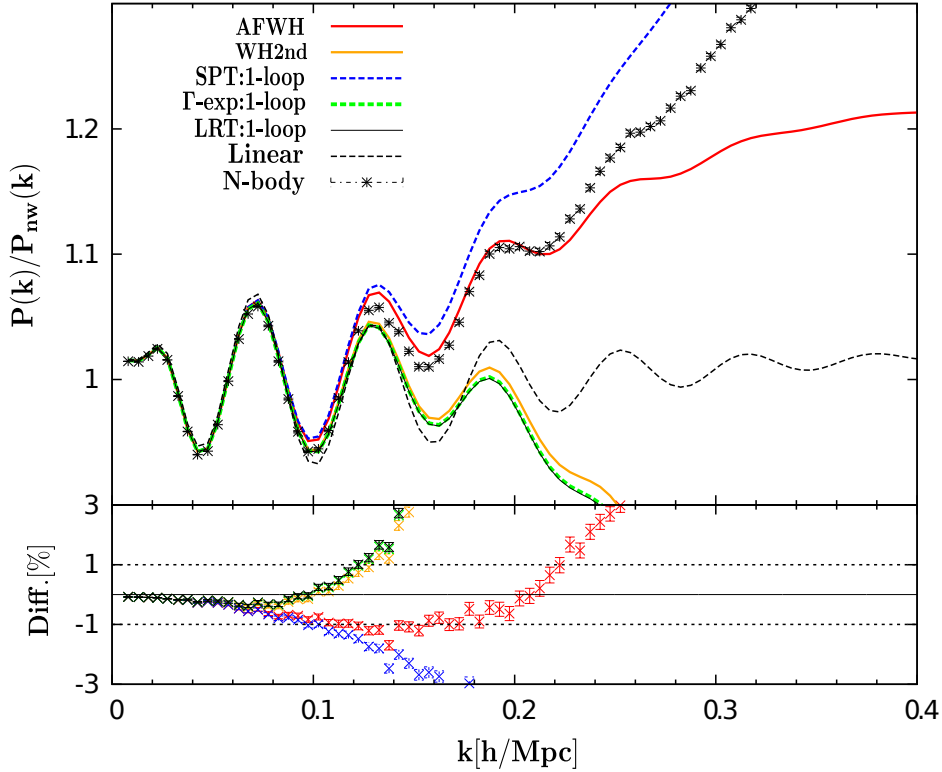}
				\end{center}
\caption{ 
          Comparison between $N$-body results and some analytical predictions in the case of WMAP 5 year cosmological parameters.
		  The results at redshifts $z = 1$ up to $k = 0.4\ h{\rm Mpc^{-1}}$ are shown.
          We show the ratio of the predicted power spectra to the smoothed reference spectra, $P(k)/P_{\rm nw}(k)$,
		  (Blue dashed, green dashed, black solid, orange solid, red solid lines, and black symbols are, respectively,
		  1-loop SPT, Regularized $\Gamma$-expansion, LRT, 2nd order of WH expansion in Eq.~(\ref{wh12})
		  and AFWH in Eq.~(\ref{result1}) predictions 
		  and $N$-body simulation result.),
		  and the fractional difference between $N$-body and analytic predicted results, $[P_{\rm Nbody}(k)-P(k)]/P_{\rm nw}(k)$,
		  (Blue, green, orange, black and red symbols are the fractional difference between $N$-body 
		  and 1-loop, Regularized $\Gamma$-expansion, 2nd order of WH expansion, LRT, and AFWH. ).
				 }
		\label{fig:1loop}
\end{figure}

In Fig.~\ref{fig:1loop},
we plot the various analytic solutions with the 1-loop level corrections and $N$-body simulation result
(blue dashed: SPT, green dashed: Regularized $\Gamma$-expansion, 
black solid: LRT, orange solid: $P_{\rm WH}^{(1)} + P_{\rm WH}^{(2)}$ in Eq.~(\ref{wh12});
red solid: AFWH in Eq.~(\ref{result1}); and black symbols: $N$-body result) at $z=1$.
\footnote{The power spectra of SPT and LRT~\citep{Matsubara:2007wj} are given by
\begin{equation}
		P_{\rm 1loop} = P_L + P_{13} + P_{22},
		\label{}
\end{equation}
\begin{equation}
		P_{\rm Lag} = \exp\left( -k^2 \sigma_v^2 \right) \left( P_L + P_{13} + P_{22} + k^2 \sigma_v^2 P_L \right),
		\label{Lag}
\end{equation}
where we denote as $P_{13} = 2 \delta_L \delta_3^{(1)}$.
Although LRT is very similar to the regularized $\Gamma$-expansion
and our result, the complete correspondence (e.g., the origin of the exponential factor) is not trivial.
}
To easily see the BAO, 
we plot the ratio of power spectrum to a smooth reference spectrum, $P(k)/P_{\rm nw}(k)$,
where the function $P_{\rm nw}(k)$ is the linear power spectrum calculated from 
the smoothed transfer function neglecting the BAO feature in \citep{Eisenstein:1997ik}.
To investigate the agreement with $N$-body results in more quantitative ways,
we also plot the fractional differences between $N$-body simulations and the predicted power spectrum $P(k)$
, i.e., $[P_{\rm Nbody}(k)-P(k)]/P(k)$
(blue: $N$-body results versus 1-loop SPT; green: regularized $\Gamma$-expansion;
black: LRT; orange: $P_{\rm WH}^{(1)} + P_{\rm WH}^{(2)}$; red: AFWH in Eq.~(\ref{result1})).

The regularized $\Gamma$-expansion,
LRT, and $P_{\rm WH}^{(1)} + P_{\rm WH}^{(2)}$ are very similar that 
we can hardly see any difference.
Their solutions improve the overestimation of SPT, but decay at low $k$ soon because of their exponential factor.
On the other hand, we can find that the main difference of our result from the previous works with 1-loop level is 
the higher order of the WH expansion, $P_{\rm WH}^{(r+1)} (r \geq 2)$.
Because of these terms,
the AFWH in Eq.~(\ref{result1}) (red solid line) does not decay and keeps the values well around those from $N$-body simulations on BAO scales.

\subsection{Comparison with 2-loop solutions in SPT}

One merit of our interpolation is that 
we can directly compare our approximate solutions with ones of each order in the perturbation theory.
In previous works, the validity of the predicted power spectra has been verified only by comparing with the $N$-body results.
However, we can verify the validity of our approximations, such as Eq.~(\ref{h1}) and Eq.~(\ref{WH r>1})
by comparing with the 2-loop solutions in SPT.
The 2-loop corrections are given by
\begin{equation}
		P_{\rm 2loop} = P_{15} + P_{24} + P_{33} + \left[ \delta_3^{(1)} \right]^2,
		\label{}
\end{equation}
where each term is calculated, respectively, as
\begin{eqnarray}
		\nonumber
		P_{15}(z,k) &=& 30 P_L(z,k) \int \frac{d^3p_1}{(2\pi)^3}\frac{d^3p_2}{(2\pi)^3} F_5(\kk,\pp_1,-\pp_1,\pp_2,-\pp_2) P_L(z,p_1) P_L(z,p_2), \\
		\nonumber
		P_{24}(z,k) &=& 24 \int \frac{d^3p_1}{(2\pi)^3} \frac{d^3p_2}{(2\pi)^3} 
		          F_2(\kk-\pp_1,\pp_1) F_4(\kk-\pp_1,\pp_1,\pp_2,-\pp_2) P_L(z,|\kk-\pp_1|) P_L(z,p_1) P_L(z,p_2),\\
		P_{33}(z,k) &=& 6 \int \frac{d^3p_1}{(2\pi)^3} \frac{d^3p_2}{(2\pi)^3} \left[ F_3(\kk-\pp_1-\pp_2,\pp_1,\pp_2) \right]^2
            	   P_L(z,|\kk-\pp_1-\pp_2|) P_L(z,p_1) P_L(z,p_2).
		\label{2-loop}
\end{eqnarray}
On the other hand, we show the corresponding approximate solutions using Eq.~(\ref{wh1ap}) and Eq.~(\ref{whrap}) as follows,
\begin{eqnarray}
		\nonumber
		P_{15} &\to& \left[ P_{15} \right]_{\rm ap} =  \frac{1}{2}\left( - \frac{k^2 \sigma_v^2}{2} \right) P_{13},  \\
		\nonumber
		P_{24} &\to& \left[ P_{24} \right]_{\rm ap} = -(k^2 \sigma_v^2)P_{22} , \\
		P_{33} &\to& \left[ P_{33} \right]_{\rm ap} =  \left( \frac{k^2 \sigma_v^2}{2} \right) P_{22},\\
		P_{\rm 2loop} &\to&\left[ P_{\rm 2loop} \right]_{\rm ap} 
		=- \frac{k^2 \sigma_v^2}{4}P_{13} - \frac{k^2 \sigma_v^2}{2}P_{22} +\left[ \delta_3^{(1)} \right]^2. 
		\label{2-loop apro}
\end{eqnarray}
Here, we have not considered the approximation of the term $\left[ \delta_3^{(1)} \right]^2$,
because it is the square of the 1loop term and we can easily compute it.
\begin{figure}[!t]
		\begin{tabular}{cc}
				\begin{minipage}{0.5\hsize}
				\begin{center}
						\epsscale{0.90}
						\plotone{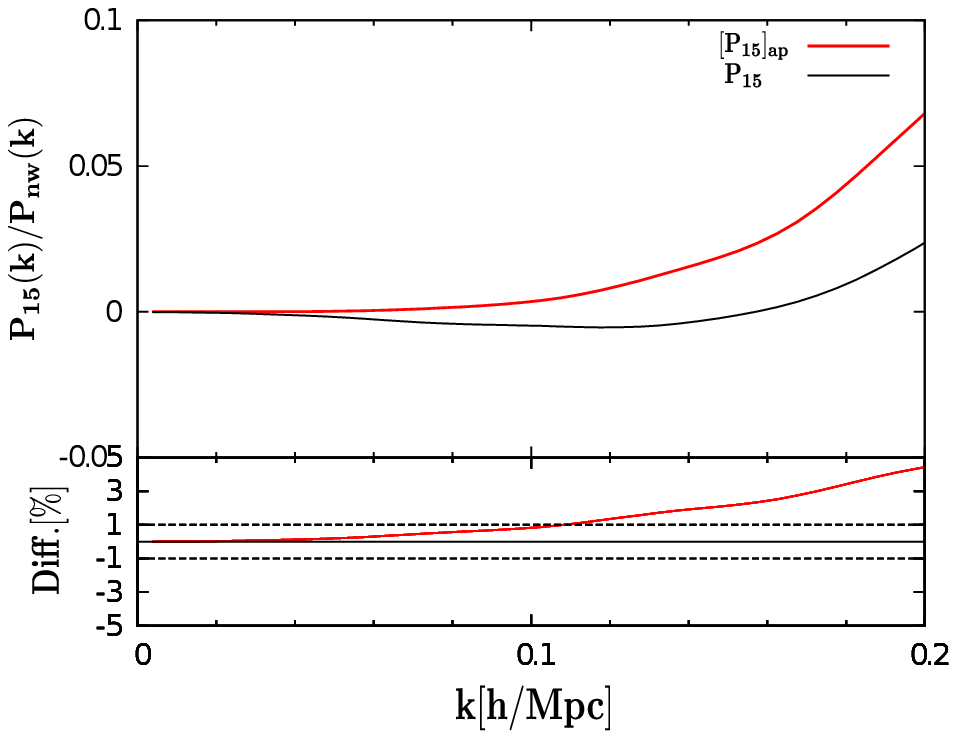}
				\end{center}
		\end{minipage}
		\begin{minipage}{0.5\hsize}
				\begin{center}
						\epsscale{0.90}
						\plotone{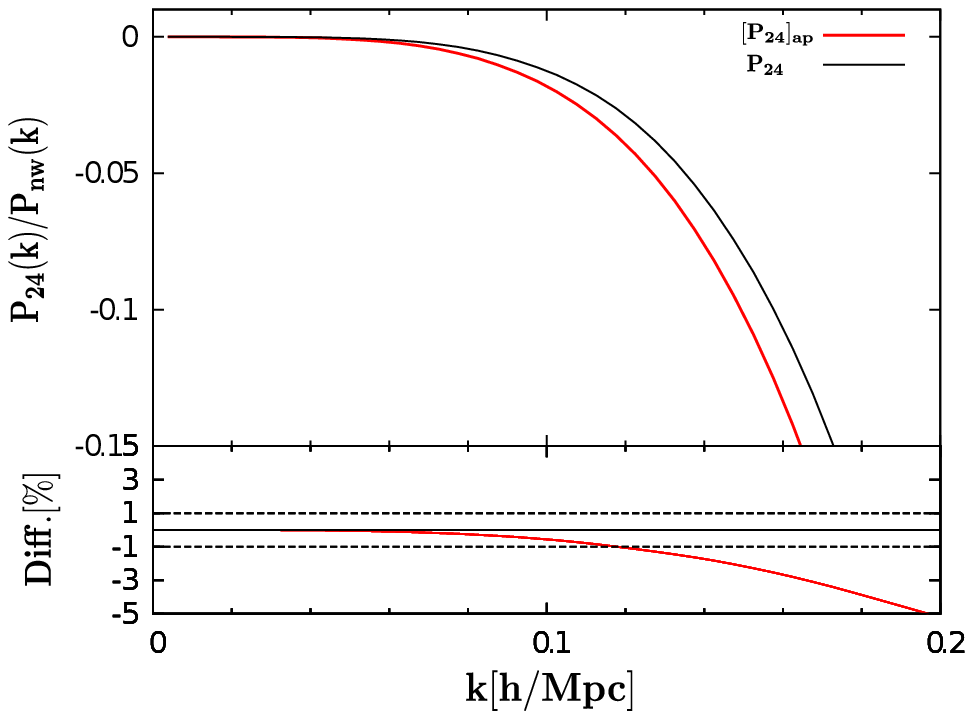}
				\end{center}
		\end{minipage}
\\
\begin{minipage}{0.5\hsize}
				\begin{center}
						\epsscale{0.90}
						\plotone{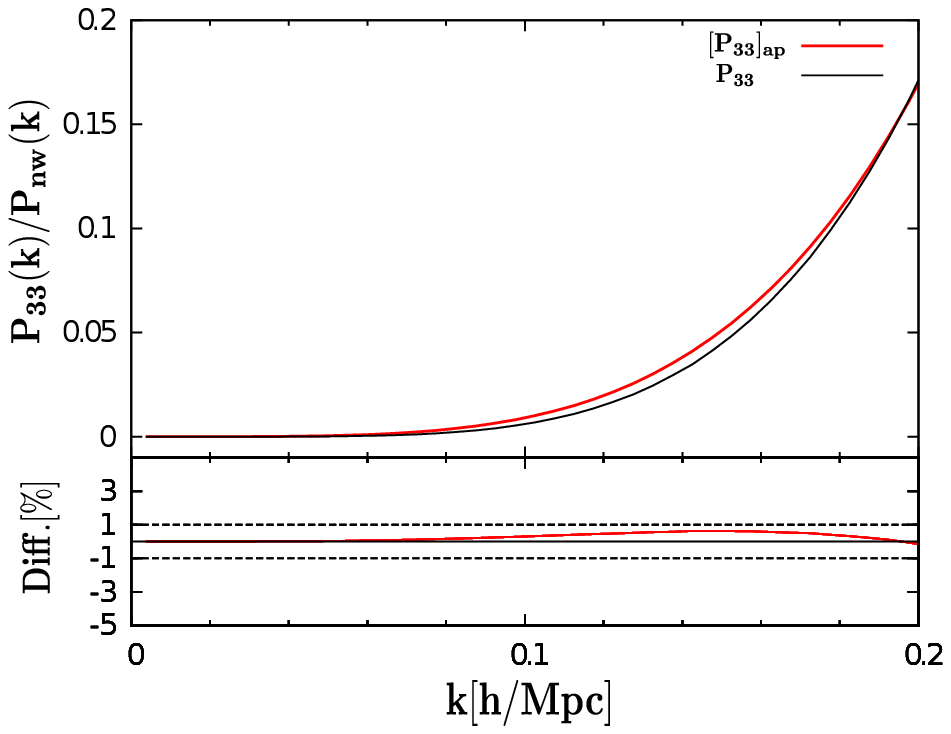}
				\end{center}
		\end{minipage}
		\begin{minipage}{0.5\hsize}
				\begin{center}
						\epsscale{0.90}
						\plotone{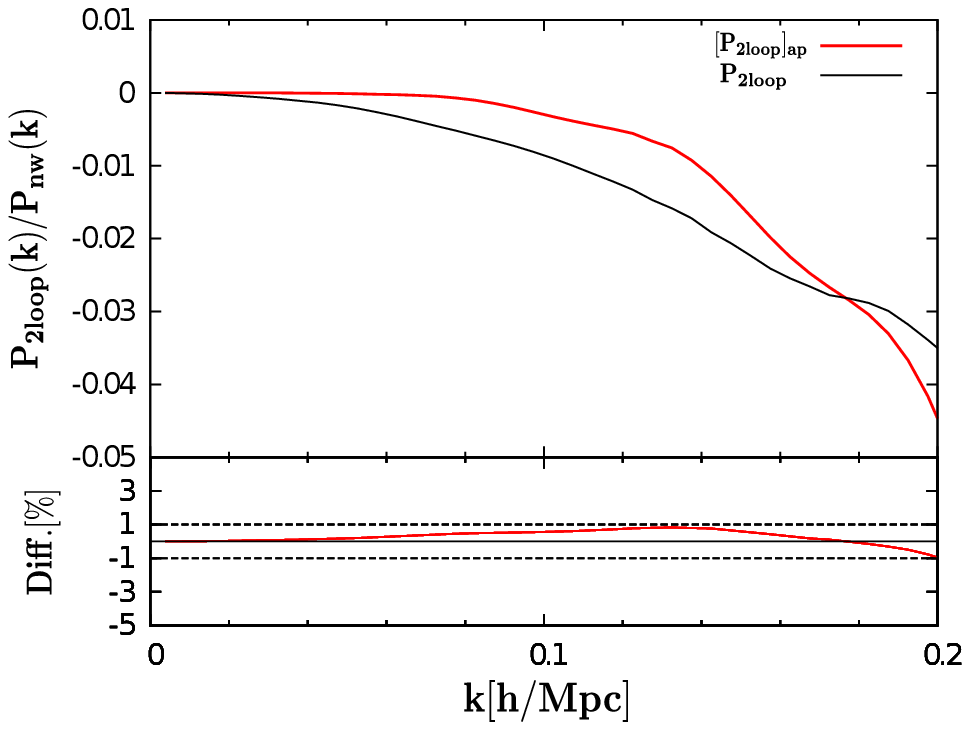}
				\end{center}
		\end{minipage}
\end{tabular}
\caption{Comparison between the approximate solutions and precise ones in the SPT 2-loop level
         in the case of WMAP 5 year cosmological parameters at $z = 1$.
		 We show the ratio of the power spectra to the smoothed reference spectra, $P/P_{\rm nw}$, $[P]_{\rm ap}/P_{\rm nw}$,
		 and the fractional difference, $[P_{\rm ap} - P]/P_{\rm nw}$ 
		 (top left: $P = P_{15}$, top right: $P_{24}$, bottom left: $P_{33}$, bottom right: $P_{\rm 2loop}$).}
		\label{fig:2loop}
\end{figure}

In Fig.~\ref{fig:2loop},
we plot the correct 2-loop solutions, their approximate solutions,
and their fractional difference, $[ [P]_{\rm ap} - P]/P_{\rm nw}$, ($P = P_{15}$, $P_{24}$, $P_{33}$, and $P_{\rm 2loop}$), in each panel.
We plot the solutions up to $0.2 h{\rm Mpc}^{-1}$ at $z=1$,
because the 2-loop corrections give a good result up to about these scales (see~\citep{Okamura:2011nu}).

For $P_{15}$ and $P_{24}$ in the top panels, the approximate solutions
respectively, are overestimated and underestimated by about 5\% at $k = 0.2 h {\rm Mpc}^{-1}$.
On the other hand, for $P_{33}$ in the bottom left panel, the approximate solution
coincides very well with the precise solution within 1\%.
One may think that since there is a large difference between $P_{15}$ and $P_{24}$, our approximation is not valid.
However, remember that each correction term in the perturbation theory tends to cancel out, resulting in small corrections.
Therefore, if the approximate solution for $P_{15}$ is overestimated, 
it would be natural that there is an underestimation in other solutions such as $P_{24}$ to cancel out the overestimated solutions.
As a result, for the full 2-loop corrections in the bottom right panel,
the fractional difference becomes within 1\% up to $0.2 h {\rm Mpc}^{-1}$.

\subsection{Comparison with closure theory}
\label{closure}
Finally, we compare with the closure theory (second Born) in \citep{Taruya:2009ir}, which is one of the best predictions at the moment.
In addition, we plot high-resolution $N$-body simulations presented by~\citep{Valageas:2011up}.

\begin{figure}[!t]
		\begin{center}
				\epsscale{1.0}
				\plotone{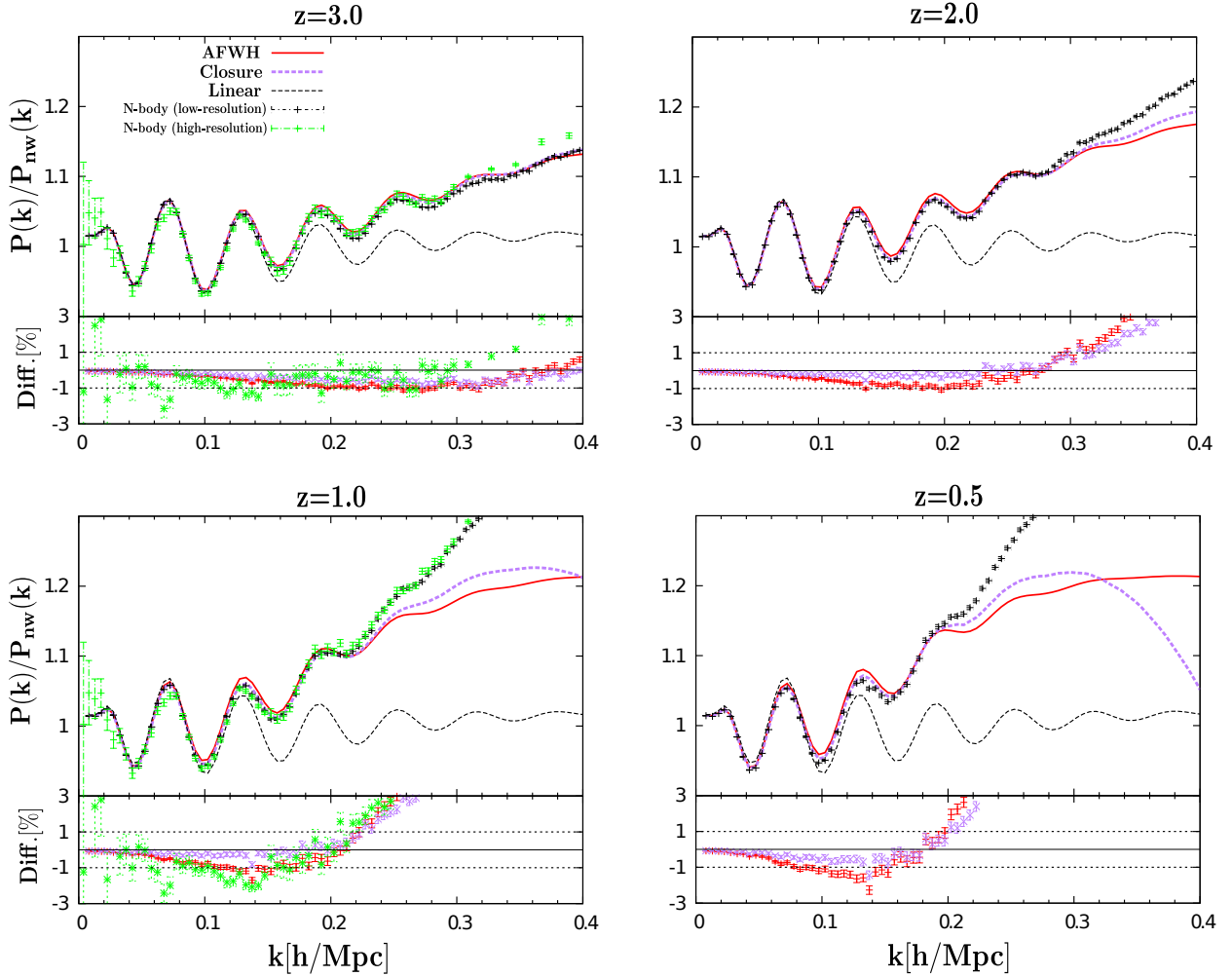}
		\end{center}
		\caption{
		This figure is the same as in Fig.~\ref{fig:1loop},
		but here we compare the predictions of the AFWH (red solid) with those of the closure theory (purple dashed)
		and $N$-body simulation results (green: high-resolution, black: low-resolution)
		at some redshifts ($z=0.5,\ 1,\ 2,\ 3$).
		We also the fractional difference between the predicted power spectra and $N$-body results.
		The red, green and purple symbols are respectively
		$N$-body (low-resolution) vs. AFWH,  $N$-body (high-resolution) vs. AFWH, 
		and $N$-body (low-resolution) vs. closure theory.}
		\label{fig:closure}
\end{figure}

In Fig.~\ref{fig:closure}, 
we plot the power spectra from closure theory (purple dashed), AFWH in Eq.~(\ref{result1}) (red solid), and $N$-body results with 
error bars (black symbols: low resolution; green symbols: high resolution) at some redshifts ($z=3.0$, $2.0$, $1.0$, $0.5$).
The range of plotted scales is $k\leq 0.4 {\rm h Mpc^{-1}}$.
We also plot the fractional differences
(purple: $N$-body result with low resolution versus closure; red: AFWH;
green: $N$-body result with high-resolution versus AFWH)

Overall, 
the predictions of AFWH tend to overestimate the $N$-body simulations at low-$k$ (BAO scales) slightly, 
and then begin to underestimate at high $k$.
The overestimation is due to the fact that  
we computed  only the 1-loop level in SPT precisely.
In fact, our approximate solutions are slightly larger than the 2-loop SPT solutions as shown Fig.~\ref{fig:2loop}.
Therefore, to derive more precise prediction on BAO scales,
we need to calculate up to the 2-loop level corrections.
The reason for the underestimation is that
either the expression for the high $k$ limit would not apply perfectly to the range of calculation 
or the subleading contributions on small scales would become effective.
We would need to compute the higher order of the WH expansion without the approximation to derive the precise nonlinearity in the high-$k$ range.
Although our results certainly give slightly less accuracy 
than that of closure theory, the difference is controlled within 1\% on BAO scales.

\section{ Correlation function}

\begin{figure}[!t]
		\begin{center}
				\epsscale{1.0}
				\plotone{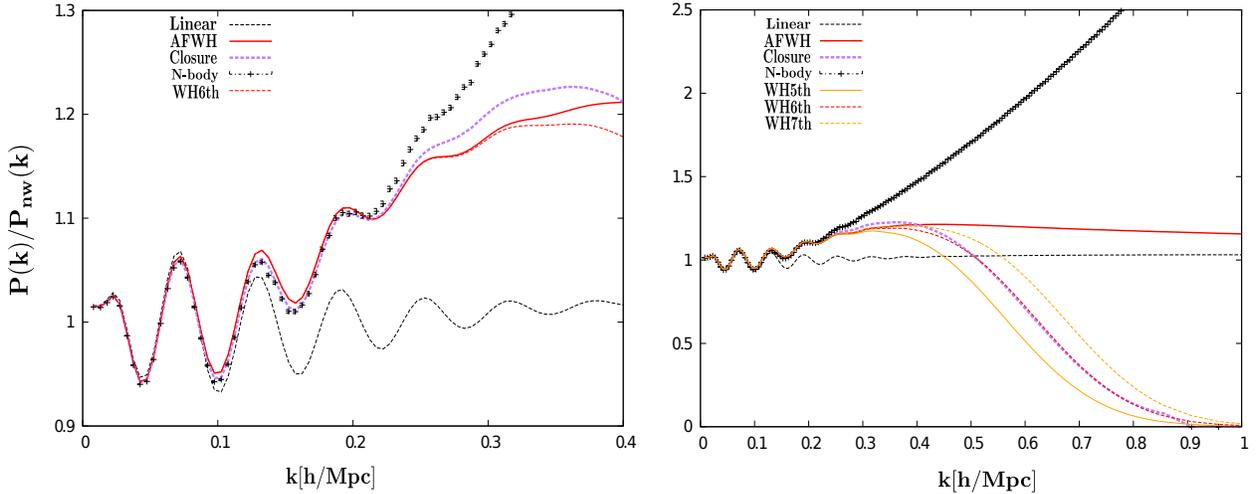}
		\end{center}
		\caption{The left and right panels are the same as the botton left panel in Fig.~\ref{fig:closure}.
		We compare the finite truncation of the WH expansion in Eq.~(\ref{WH6th}) (red dashed) 
		with the AFWH in Eq.~(\ref{result1}) (red line) at $z=1$.
		In the right panel, we further plot the 5th, 6th and 7th order of the WH expansion up to $k=1[h/{\rm Mpc}]$.}
		\label{fig:AFPT}
\end{figure}

\begin{figure}[!t]
		\begin{center}
				\epsscale{1.0}
				\plotone{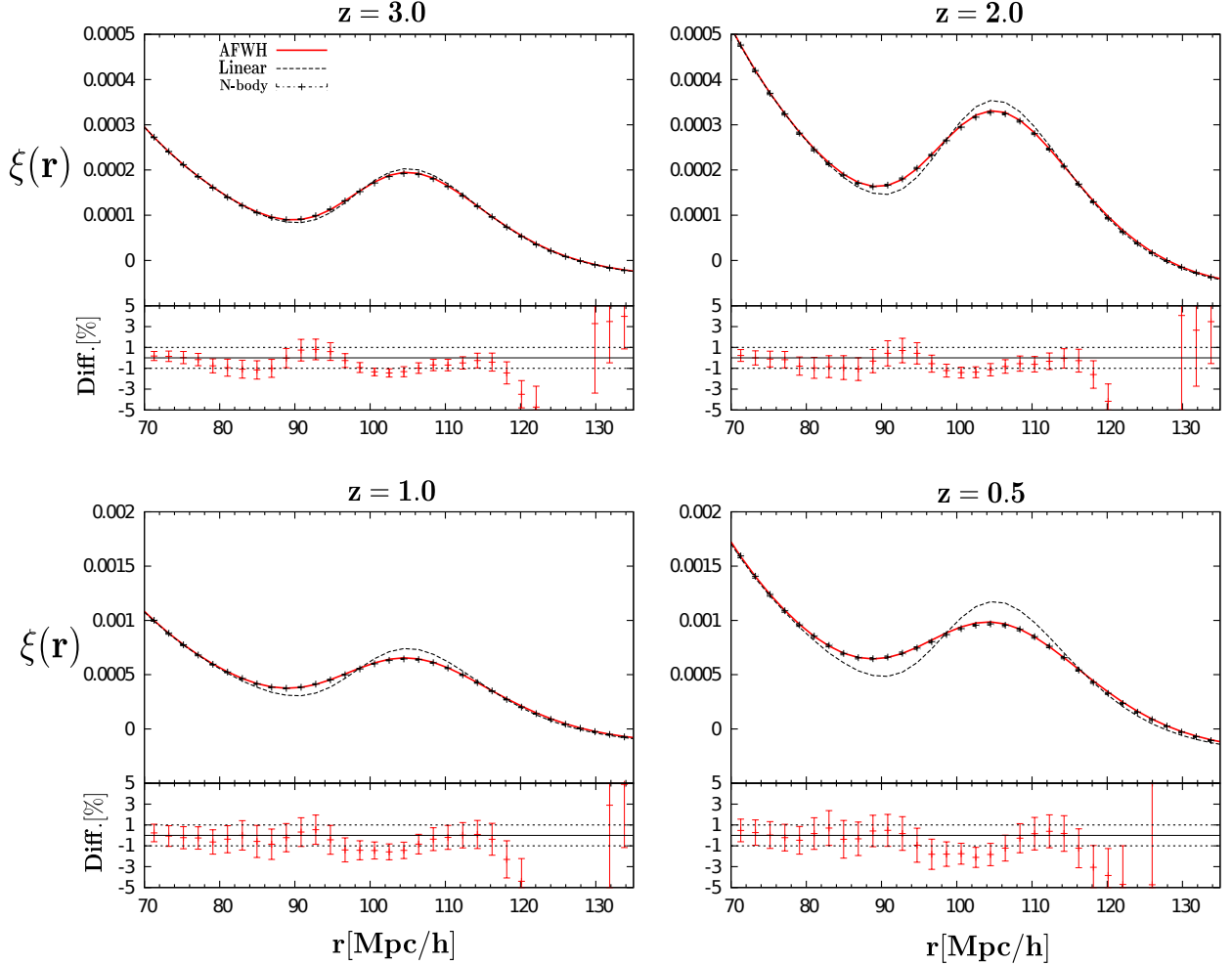}
		\end{center}
		\caption{Comparison between the predicted correlation functions and $N$-body results
		(red line: AFWH, black dashed: linear theory, and black symbols: $N$-body simulations).
		The results at some redshits ($z$=0.5, 1.0, 2.0, 3.0) with the range of $70 \leq r [{\rm Mpc}/h]\leq135 $ are shown. 
		We further plot the fractional difference between the predictions of the AFWH and $N$-body results, 
		$[\xi_{\rm Nbody}(r)-\xi(r)]/\xi(r)$.}
		\label{fig:correlation}
\end{figure}

We compute the two-point correlation function calculated from the power spectrum in Eq.~(\ref{result1}),
which is given by
\begin{equation}
		\xi(z,r) = \int_0^{\infty} \frac{k^2 dk}{2 \pi^2} \frac{\sin(kr)}{kr} P_{\rm AF}(z,k).
		\label{correlation}
\end{equation}

Usually, we are not able to compute the correlation function in SPT, because the integrand function diverge.
For example, in the 1-loop level of SPT, the predicted solution has the scale dependence of $k^2 P_L(k)$
at high $k$, because the approximate solutions of $P_{13}$ and $P_{22}$ are proportional to $k^2 P_L(k)$.
Therefore, the integrand in Eq.~(\ref{correlation}) has the scale-dependence of $\sin(kr)\ln^2(k)$:
\begin{equation}
		\frac{k^2}{2 \pi^2} \frac{\sin(kr)}{kr} P_{\rm 1loop}(z,k)
		\propto \frac{k^2}{2 \pi^2} \frac{\sin(kr)}{kr} k^2 P_L(k)
		\propto \sin(kr) \ln^2(k),
		\label{}
\end{equation}
where we have used the behavior of the linear power spectrum at high-$k$, $P_L(k)\to \ln^2(k)/k^3$.
This solution diverge at high $k$, and we are not able to evaluate the integration in the range of $0\leq k \leq \infty$.

On the other hand, we are able to compute the correlation function for AFWH because the solution has the
scale dependence like the linear power spectrum at high $k$:
\begin{align}
		P_{\rm AF}(z,k) &\to \left[ \delta_L(z,k) + \frac{2\delta_3^{(1)}(z,k)}{k^2 \sigma_v^2} \right]^2
		+ \frac{P_{22}(z,k)}{k^2 \sigma_v^2}, \notag \\
		& \propto P_L(z,k).
		\label{}
\end{align}
In the first line, we dropped the terms including exponential factor.
The scale-dependence of $\delta_3^{(1)}$ and $P_{22}$ at high-$k$ are proportional to $k^2 \delta_L(k)$ and $k^2 P_L(k)$.
As a result, $P_{\rm AF}$ (red line in the right panel of Fig.~\ref{fig:AFPT})
is proportional to $P_L$ (black dashed in Fig.~\ref{fig:AFPT}),
and the integrand of Eq.~(\ref{correlation}) converge like the linear power spectrum.

Furthermore, as long as we focus on the BAO scales in real space ($60 \lesssim r \lesssim 140 [{\rm Mpc}/h]$),
the behavior of the power spectrum on small scales in Fourier space ($k \geq 0.2{\rm -}0.4 [h/{\rm Mpc}]$)
contributes very little to the result of the correlation function.
Therefore, we may truncate the WH expansion up to the appropriate order 
so that the integrand function converge to zero due to the exponential factor and we can easily compute the correlation function:
\begin{align}
		P_{\rm AF}(z,k)	  \to & \left[ \delta_L(z,k) - \frac{2\delta_3^{(1)}(z,k)}{k^2 \sigma_v^2}
				 \left( \exp\left(-\frac{k^2 \sigma_v^2}{2}\right) -1 \right) \right]^2 \notag \\
				 & + \exp(-k^2 \sigma_v^2) P_{22}\left( 1 + \frac{1}{2}(k^2 \sigma_v^2) + \frac{1}{3!} (k^2 \sigma_v^2)^2 + 
				 \frac{1}{4!} (k^2 \sigma_v^2)^3 +\frac{1}{5!} (k^2 \sigma_v^2)^4\right),
		\label{WH6th}
\end{align}
where we have truncated the WH expansion up to the sixth order.
We plot the solution of Eq.~(\ref{WH6th}) in Fig.~\ref{fig:AFPT} at $z=1$,
where the difference between the AFWH in Eq.~(\ref{result1}) (red line) and the solution of Eq.~(\ref{WH6th}) 
(red dashed) up to $k \lesssim0.25[h/{\rm Mpc}]$ is not visible,
and the solution behaves like the ones of closure theory at high-$k$.
In the right panel of Fig.~\ref{fig:AFPT},
we further plot the solutions of the fifth (orange line) and seventh (orange dashed) order in the WH expansion.

Here, we adopt the sixth order solution of the WH expansion to compute the correlation function in Eq.~(\ref{correlation}).
In Fig.~\ref{fig:correlation},
we plot the analytic predictions of the correlation function, $\xi(r)$
(red line: AFWH; black dashed: linear theory; black symbols: $N$-body results),
and the fractional difference between the predicted correlation functions from AFWH and the $N$-body simulation results,
$[\xi_{\rm Nbody}(r)-\xi(r)]/\xi(r)$.

Our predictions explain the displacement of the location of the BAO peaks 
and the smoothing of their amplitudes due to the non-linear effects.
As a result, the fractional difference against the $N$-body results is within 1--3\%.
Almost the same results are derived even for the second order of the WH expansion in Eq.~(\ref{wh12}),
because on small scales there is very little contribution to the correlation function around the BAO peak.
This fact is also well known also in other modified perturbation theories (e.g., \citep{Taruya:2009ir,Okamura:2011nu}).

\section{Conclusion}
We have applied the WH expansion to the evolution equation of dark matter in Newtonian gravity. 
It diagrammatically corresponds to the classification of the power spectrum in which each order includes all of the vertex loop contributions.
It is proved that the WH expansion is mathematically equivalent to the $\Gamma$-expansion approach in the Multi-Point Propagators method. 

Even if WH expansion method is physically and mathematically useful for understanding the non-linearity of the evolution of dark matter, 
the validity of the finite truncation of the expansion is not clear 
and the difficulty in the calculation will remain.
To resolve these difficulties, we proposed a way to include the effect of all orders by assuming that the high non-linear solutions are well approximated by the ones in the high-$k$ limit.  Namely we calculate only low order terms precisely
and replace the high order solutions with the ones in the high-$k$ limit.

It has been known in RPT
that the matter density and velocity fluctuations of dark matter are exponential in the high-$k$ limit.
We proved again this behavior in the context of SPT using the WH expansion
by proving that the kernel functions $F$ and $G$ take the form  of Eq.~(\ref{theorem}) in the high-$k$ limit.
Using the approximate kernel functions $F$ and $G$,
we proposed an appropriate interpolation between high-$k$ and low-$k$ solutions,
and the approximate full power spectrum in Eq.~(\ref{result1}), which  approximately include the full order of SPT.

We compared our results with some other analytic predictions (e.g., regularized $\Gamma$-expansion, LRT, SPT, and closure theory) 
and $N$-body simulation results.
Since the WH expansion is equivalent to the $\Gamma$-expansion and the regularized $\Gamma$-expansion bases on the $\Gamma$-expansion,
we can describe the first order of the WH expansion, $P_{\rm WH}^{(1)}$, using the regularized $\Gamma$-expansion in Eq.~(\ref{Hreg1}).
One of the difference between our result and the regularized $\Gamma$-expansion
is the manner of interpolating between the high-$k$ and low-$k$ solutions,
but this difference slightly affects the predicted power spectrum. 
Another difference is that we consider the higher order of the WH expansion approximately.
As a result, even for the 1-loop level,
the predicted power spectrum in Eq.~(\ref{result1}) does not decay due to the exponential factor as shown in Fig.~\ref{fig:1loop},
and results in good agreement with the $N$-body simulation on BAO scales.

The validity of the various modified perturbation theory (e.g., LRT, RPT, closure theory, \dots ) predictions
is usually verified only by comparing with the $N$-body simulations.
However, we can also verify our approximation by comparing with the solutions with the SPT 2loop level.
In Fig.\ref{fig:2loop}, we showed that the fractional difference between the approximate solutions and the precise solutions with
the SPT 2-loop level is within 1 \% on BAO scales ($\leq 0.2 h{\rm Mpc}^{-1}$) for the $WMAP$ five year cosmological parameters at $z=1$.

We also compared with the closure theory which is one of the best prediction at a moment,
and the accuracy of AFWH in Eq.~(\ref{result1}) is comparable to or slightly less than the ones in the closure theory,
with the fractional difference within 1\% on BAO scales.

Finally, we computed the two-point correlation function for AFWH.
We can compute the correlation functions because the predicted power spectrum in AFWH converges like the one from linear theory.
Since the contributions on small scales do not affect the values of the correlation function,
one may use Eq.~(\ref{WH6th}) to compute the correlation function.
This solution has the same behavior as Eq.~(\ref{result1}) on BAO scales and decay on small scales due to the exponential factor
in Fig.~\ref{fig:AFPT}.
The predicted correlation functions from the AFWH agree very well with the $N$-body simulation results,
and the fractional difference is within $1-3$\%.

We could use and apply our results to various studies of the nonlinear evolution of dark matter
(e.g., redshift distortion effect, bias effect, and bispectrum, etc.),
because our prescription is easy and gives good results that are comparable to closure theory, 
and furthermore the computational time is very rapid.

\section*{Acknowledgments}
We would like to thank T. Nishimichi and A. Taruya for providing us with the numerical simulation results and useful comments
and Y. Itoh, and T. Okamura for useful discussion. This work is supported in part by the GCOE Program ``Weaving Science Web beyond
Particle-matter Hierarchy'' at Tohoku University and
by a Grant-in-Aid for Scientific Research from JSPS 
(No. 24-3849 for NSS and Nos. 18072001, 20540245 for TF) as well as by Core-to-Core Program
``International Research Network for Dark Energy.''
TF thanks Luc Branchet and the Institute of Astronomical Observatory, Paris for warm hospitality during his stay in the last stage 
of the present work.

\bibliographystyle{JHEP}

\begin{thebibliography}{10}

\bibitem{Eisenstein/etal:1998tu}
D.~J. Eisenstein, W.~Hu, and M.~Tegmark, {\it {Cosmic complementarity: H(0) and
  Omega(m) from combining CMB experiments and redshift surveys}},  {\em
  Astrophys. J.} {\bf 504} (1998) L57--L61,
  [\href{http://xxx.lanl.gov/abs/astro-ph/9805239}{{\tt astro-ph/9805239}}].

\bibitem{Matsubara:2004fr}
T.~Matsubara, {\it {Correlation Function in Deep Redshift Space as a
  Cosmological Probe}},  {\em Astrophys. J.} {\bf 615} (2004) 573--585,
  [\href{http://xxx.lanl.gov/abs/astro-ph/0408349}{{\tt astro-ph/0408349}}].

\bibitem{Eisenstein/etal:2005}
{\bf SDSS} Collaboration, D.~J. Eisenstein {\em et.~al.}, {\it {Detection of
  the Baryon Acoustic Peak in the Large-Scale Correlation Function of SDSS
  Luminous Red Galaxies}},  {\em Astrophys. J.} {\bf 633} (2005) 560--574,
  [\href{http://xxx.lanl.gov/abs/astro-ph/0501171}{{\tt astro-ph/0501171}}].

\bibitem{Seo:2003pu}
H.-J. Seo and D.~J. Eisenstein, {\it {Probing dark energy with baryonic
  acoustic oscillations from future large galaxy redshift surveys}},  {\em
  Astrophys.J.} {\bf 598} (2003) 720--740,
  [\href{http://xxx.lanl.gov/abs/astro-ph/0307460}{{\tt astro-ph/0307460}}].

\bibitem{Blake:2003rh}
C.~Blake and K.~Glazebrook, {\it {Probing dark energy using baryonic
  oscillations in the galaxy power spectrum as a cosmological ruler}},  {\em
  Astrophys.J.} {\bf 594} (2003) 665--673,
  [\href{http://xxx.lanl.gov/abs/astro-ph/0301632}{{\tt astro-ph/0301632}}].

\bibitem{Glazebrook:2005mb}
K.~Glazebrook and C.~Blake, {\it {Measuring the cosmic evolution of dark energy
  with baryonic oscillations in the galaxy power spectrum}},  {\em
  Astrophys.J.} {\bf 631} (2005) 1--20,
  [\href{http://xxx.lanl.gov/abs/astro-ph/0505608}{{\tt astro-ph/0505608}}].

\bibitem{Shoji:2008xn}
M.~Shoji, D.~Jeong, and E.~Komatsu, {\it {Extracting Angular Diameter Distance
  and Expansion Rate of the Universe from Two-dimensional Galaxy Power Spectrum
  at High Redshifts: Baryon Acoustic Oscillation Fitting versus Full
  Modeling}},  {\em Astrophys. J.} {\bf 693} (2009) 1404--1416,
  [\href{http://xxx.lanl.gov/abs/0805.4238}{{\tt arXiv:0805.4238}}].

\bibitem{Padmanabhan:2008ag}
N.~Padmanabhan and .~White, Martin~J., {\it {Constraining Anisotropic Baryon
  Oscillations}},  {\em Phys. Rev.} {\bf D77} (2008) 123540,
  [\href{http://xxx.lanl.gov/abs/0804.0799}{{\tt arXiv:0804.0799}}].

\bibitem{Fry:1983cj}
J.~N. Fry, {\it {The Galaxy correlation hierarchy in perturbation theory}},
  {\em Astrophys. J.} {\bf 279} (1984) 499--510.

\bibitem{Goroff:1986ep}
M.~H. Goroff, B.~Grinstein, S.~J. Rey, and M.~B. Wise, {\it {Coupling of Modes
  of Cosmological Mass Density Fluctuations}},  {\em Astrophys. J.} {\bf 311}
  (1986) 6--14.

\bibitem{Suto:1990wf}
Y.~Suto and M.~Sasaki, {\it {Quasi nonlinear theory of cosmological
  selfgravitating systems}},  {\em Phys. Rev. Lett.} {\bf 66} (1991) 264--267.

\bibitem{Makino:1991rp}
N.~Makino, M.~Sasaki, and Y.~Suto, {\it {Analytic approach to the perturbative
  expansion of nonlinear gravitational fluctuations in cosmological density and
  velocity fields}},  {\em Phys. Rev.} {\bf D46} (1992) 585--602.

\bibitem{Jain:1993jh}
B.~Jain and E.~Bertschinger, {\it {Second order power spectrum and nonlinear
  evolution at high redshift}},  {\em Astrophys. J.} {\bf 431} (1994) 495,
  [\href{http://xxx.lanl.gov/abs/astro-ph/9311070}{{\tt astro-ph/9311070}}].

\bibitem{Scoccimarro:1996se}
R.~Scoccimarro and J.~Frieman, {\it {Loop Corrections in Non-Linear
  Cosmological Perturbation Theory II. Two-point Statistics and
  Self-Similarity}},  {\em Astrophys. J.} {\bf 473} (1996) 620,
  [\href{http://xxx.lanl.gov/abs/astro-ph/9602070}{{\tt astro-ph/9602070}}].

\bibitem{Bernardeau:2001qr}
F.~Bernardeau, S.~Colombi, E.~Gaztanaga, and R.~Scoccimarro, {\it {Large-scale
  structure of the universe and cosmological perturbation theory}},  {\em Phys.
  Rept.} {\bf 367} (2002) 1--248,
  [\href{http://xxx.lanl.gov/abs/astro-ph/0112551}{{\tt astro-ph/0112551}}].

\bibitem{Jeong:2006xd}
D.~Jeong and E.~Komatsu, {\it {Perturbation Theory Reloaded: Analytical
  Calculation of Non-linearity in Baryonic Oscillations in the Real Space
  Matter Power Spectrum}},  {\em Astrophys. J.} {\bf 651} (2006) 619--626,
  [\href{http://xxx.lanl.gov/abs/astro-ph/0604075}{{\tt astro-ph/0604075}}].

\bibitem{Jeong:2008rj}
D.~Jeong and E.~Komatsu, {\it {Perturbation Theory Reloaded II: Non-linear
  Bias, Baryon Acoustic Oscillations and Millennium Simulation In Real Space}},
   {\em Astrophys. J.} {\bf 691} (2009) 569--595,
  [\href{http://xxx.lanl.gov/abs/0805.2632}{{\tt arXiv:0805.2632}}].

\bibitem{Crocce:2005xy}
M.~Crocce and R.~Scoccimarro, {\it {Renormalized Cosmological Perturbation
  Theory}},  {\em Phys. Rev.} {\bf D73} (2006) 063519,
  [\href{http://xxx.lanl.gov/abs/astro-ph/0509418}{{\tt astro-ph/0509418}}].

\bibitem{Crocce:2005xz}
M.~Crocce and R.~Scoccimarro, {\it {Memory of Initial Conditions in
  Gravitational Clustering}},  {\em Phys. Rev.} {\bf D73} (2006) 063520,
  [\href{http://xxx.lanl.gov/abs/astro-ph/0509419}{{\tt astro-ph/0509419}}].

\bibitem{Crocce:2007dt}
M.~Crocce and R.~Scoccimarro, {\it {Nonlinear Evolution of Baryon Acoustic
  Oscillations}},  {\em Phys. Rev.} {\bf D77} (2008) 023533,
  [\href{http://xxx.lanl.gov/abs/0704.2783}{{\tt arXiv:0704.2783}}].

\bibitem{Scoccimarro:1997gr}
R.~Scoccimarro, {\it {Transients from initial conditions: a perturbative
  analysis}},  {\em Mon.Not.Roy.Astron.Soc.} {\bf 299} (1998) 1097,
  [\href{http://xxx.lanl.gov/abs/astro-ph/9711187}{{\tt astro-ph/9711187}}].

\bibitem{Taruya:2009ir}
A.~Taruya, T.~Nishimichi, S.~Saito, and T.~Hiramatsu, {\it {Non-linear
  Evolution of Baryon Acoustic Oscillations from Improved Perturbation Theory
  in Real and Redshift Spaces}},  {\em Phys. Rev.} {\bf D80} (2009) 123503,
  [\href{http://xxx.lanl.gov/abs/0906.0507}{{\tt arXiv:0906.0507}}].

\bibitem{Hiramatsu:2009ki}
T.~Hiramatsu and A.~Taruya, {\it {Chasing the non-linear evolution of matter
  power spectrum with numerical resummation method: solution of closure
  equations}},  {\em Phys. Rev.} {\bf D79} (2009) 103526,
  [\href{http://xxx.lanl.gov/abs/0902.3772}{{\tt arXiv:0902.3772}}].

\bibitem{Pietroni:2008jx}
M.~Pietroni, {\it {Flowing with Time: a New Approach to Nonlinear Cosmological
  Perturbations}},  {\em JCAP} {\bf 0810} (2008) 036,
  [\href{http://xxx.lanl.gov/abs/0806.0971}{{\tt arXiv:0806.0971}}].

\bibitem{Bernardeau:2008fa}
F.~Bernardeau, M.~Crocce, and R.~Scoccimarro, {\it {Multi-Point Propagators in
  Cosmological Gravitational Instability}},  {\em Phys. Rev.} {\bf D78} (2008)
  103521, [\href{http://xxx.lanl.gov/abs/0806.2334}{{\tt arXiv:0806.2334}}].

\bibitem{Bernardeau:2010md}
F.~Bernardeau, M.~Crocce, and E.~Sefusatti, {\it {Multi-Point Propagators for
  Non-Gaussian Initial Conditions}},  {\em Phys.Rev.} {\bf D82} (2010) 083507,
  [\href{http://xxx.lanl.gov/abs/1006.4656}{{\tt arXiv:1006.4656}}].

\bibitem{Bernardeau:2011vy}
F.~Bernardeau, N.~Van~de Rijt, and F.~Vernizzi, {\it {Resummed propagators in
  multi-component cosmic fluids with the eikonal approximation}},
  \href{http://xxx.lanl.gov/abs/1109.3400}{{\tt arXiv:1109.3400}}.

\bibitem{Bernardeau:2011dp}
F.~Bernardeau, M.~Crocce, and R.~Scoccimarro, {\it {Constructing Regularized
  Cosmic Propagators}},  \href{http://xxx.lanl.gov/abs/1112.3895}{{\tt
  arXiv:1112.3895}}.

\bibitem{McDonald:2006hf}
P.~McDonald, {\it {Dark matter clustering: a simple renormalization group
  approach}},  {\em Phys. Rev.} {\bf D75} (2007) 043514,
  [\href{http://xxx.lanl.gov/abs/astro-ph/0606028}{{\tt astro-ph/0606028}}].

\bibitem{Valageas:2003gm}
P.~Valageas, {\it {A new approach to gravitational clustering: a path- integral
  formalism and large-N expansions}},  {\em Astron. Astrophys.} {\bf 421}
  (2004) 23--40, [\href{http://xxx.lanl.gov/abs/astro-ph/0307008}{{\tt
  astro-ph/0307008}}].

\bibitem{Matarrese:2007wc}
S.~Matarrese and M.~Pietroni, {\it {Resumming Cosmic Perturbations}},  {\em
  JCAP} {\bf 0706} (2007) 026,
  [\href{http://xxx.lanl.gov/abs/astro-ph/0703563}{{\tt astro-ph/0703563}}].

\bibitem{Matsubara:2007wj}
T.~Matsubara, {\it {Resumming Cosmological Perturbations via the Lagrangian
  Picture: One-loop Results in Real Space and in Redshift Space}},  {\em Phys.
  Rev.} {\bf D77} (2008) 063530, [\href{http://xxx.lanl.gov/abs/0711.2521}{{\tt
  arXiv:0711.2521}}].

\bibitem{Matsubara:2011ck}
T.~Matsubara, {\it {Nonlinear Perturbation Theory Integrated with Nonlocal
  Bias, Redshift-space Distortions, and Primordial Non- Gaussianity}},  {\em
  Phys. Rev.} {\bf D83} (2011) 083518,
  [\href{http://xxx.lanl.gov/abs/1102.4619}{{\tt arXiv:1102.4619}}].

\bibitem{Matsubara:2008wx}
T.~Matsubara, {\it {Nonlinear perturbation theory with halo bias and redshift-
  space distortions via the Lagrangian picture}},  {\em Phys. Rev.} {\bf D78}
  (2008) 083519, [\href{http://xxx.lanl.gov/abs/0807.1733}{{\tt
  arXiv:0807.1733}}].

\bibitem{Okamura:2011nu}
T.~Okamura, A.~Taruya, and T.~Matsubara, {\it {Next-to-leading resummation of
  cosmological perturbations via the Lagrangian picture: 2-loop correction in
  real and redshift spaces}},  {\em JCAP} {\bf 1108} (2011) 012,
  [\href{http://xxx.lanl.gov/abs/1105.1491}{{\tt arXiv:1105.1491}}].

\bibitem{Valageas:2011up}
P.~Valageas and T.~Nishimichi, {\it {Combining perturbation theories with halo
  models for the matter bispectrum}},  {\em Astron. Astrophys.} {\bf 532}
  (2011) A4, [\href{http://xxx.lanl.gov/abs/1102.0641}{{\tt arXiv:1102.0641}}].

\bibitem{Komatsu:2008hk}
{\bf WMAP} Collaboration, E.~Komatsu {\em et.~al.}, {\it {Five-Year Wilkinson
  Microwave Anisotropy Probe (WMAP\altaffilmark 1 ) Observations:Cosmological
  Interpretation}},  {\em Astrophys. J. Suppl.} {\bf 180} (2009) 330--376,
  [\href{http://xxx.lanl.gov/abs/0803.0547}{{\tt arXiv:0803.0547}}].

\bibitem{Springel:2005mi}
V.~Springel, {\it {The cosmological simulation code GADGET-2}},  {\em Mon. Not.
  Roy. Astron. Soc.} {\bf 364} (2005) 1105--1134,
  [\href{http://xxx.lanl.gov/abs/astro-ph/0505010}{{\tt astro-ph/0505010}}].

\bibitem{Crocce:2006ve}
M.~Crocce, S.~Pueblas, and R.~Scoccimarro, {\it {Transients from Initial
  Conditions in Cosmological Simulations}},  {\em Mon. Not. Roy. Astron. Soc.}
  {\bf 373} (2006) 369--381,
  [\href{http://xxx.lanl.gov/abs/astro-ph/0606505}{{\tt astro-ph/0606505}}].

\bibitem{Nishimichi:2008ry}
T.~Nishimichi {\em et.~al.}, {\it {Modeling Nonlinear Evolution of Baryon
  Acoustic Oscillations: Convergence Regime of N-body Simulations and Analytic
  Models}},  \href{http://xxx.lanl.gov/abs/0810.0813}{{\tt arXiv:0810.0813}}.

\bibitem{Eisenstein:1997ik}
D.~J. Eisenstein and W.~Hu, {\it {Baryonic Features in the Matter Transfer
  Function}},  {\em Astrophys. J.} {\bf 496} (1998) 605,
  [\href{http://xxx.lanl.gov/abs/astro-ph/9709112}{{\tt astro-ph/9709112}}].

\end{thebibliography}

\providecommand{\href}[2]{#2}\begingroup\raggedright\endgroup
\end{document}